\documentclass[aip,jcp,reprint,noshowkeys]{revtex4-1}
\usepackage{graphicx,float,dcolumn,bm,xcolor,microtype,multirow,amscd,amsmath,amssymb,amsfonts,physics,mhchem,pifont,wrapfig,enumitem}
\usepackage{libertine}
\usepackage[normalem]{ulem}
\usepackage[printonlyused]{acronym}
\DeclareMathSymbol{\shortminus}{\mathbin}{AMSa}{"39}
\usepackage{simpler-wick}
\usepackage{relsize}

\usepackage{hyperref}
\hypersetup{
    colorlinks=true,
    linkcolor=blue,
    filecolor=blue,      
    urlcolor=blue,	
    citecolor=blue
}

\renewcommand{\braket}[3]{\langle #1 | #2 | #3 \rangle}
\newcommand{\brkt}[2]{\langle #1 | #2 \rangle}
\renewcommand{\acomm}[2]{[#1,#2]_{_+}}
\newcommand{\up}[1]{\hspace{0pt}^{#1}}

\newcolumntype{Y}{>{\centering\arraybackslash}X}
\newcommand{\mc}{\multicolumn}

\newcommand{\cC}{\mathcal{\mathsmaller C}}
\newcommand{\cI}{\mathcal{I}}
\newcommand{\cZ}{\mathcal{Z}}

\newcommand{\Cn}{\cC_{n}}

\renewcommand{\a}{a^{\phantom{\dagger}}}
\renewcommand{\b}{b^{\vphantom{\dagger}}}

\newcommand{\adag}{a^{\dagger}}
\newcommand{\bdag}{b^{\dagger}}

\newcommand{\tb}{\tilde{b}}
\newcommand{\tbdag}{\tilde{b}^{\dagger}}

\newcommand{\C}{C}
\newcommand{\tC}{\tilde{C}}
\newcommand{\tS}{\tilde{S}}
\newcommand{\bX}{\bar{X}}
\newcommand{\bY}{\bar{Y}}
\newcommand{\bJ}{\bar{J}}
\newcommand{\bK}{\bar{K}}

\newcommand{\canon}[3]{\braket{#1}{#2}{#3}_{0}}

\newcommand{\Or}[1]{\mathcal{O}\qty(#1)}

\newcommand{\Ne}{N}
\newcommand{\Nbas}{n}
\newcommand{\Nref}{N_\text{ref}}

\newcommand{\vac}{\ket{-}}

\newcommand{\etal}{\textit{et al.}}
\newcommand{\eg}{e.g.}
\newcommand{\ie}{i.e.}
\definecolor{hughgreen}{RGB}{0, 139, 0}
\newcommand{\hugh}[1]{#1}

\makeatletter
\newcommand\niton{\mathrel{\m@th\mathpalette\canc@l\owns}}
\newcommand\canc@l[2]{{\ooalign{$\hfil#1/\mkern1mu\hfil$\crcr$#1#2$}}}
\makeatother

\newcommand{\UOX}{Physical and Theoretical Chemistry Laboratory, Department of Chemistry, University of Oxford, South Parks Road, Oxford, OX1 3QZ, U.K.}

\begin{document}

\title{Generalised Nonorthogonal Matrix Elements: Unifying  Wick's Theorem and the Slater--Condon Rules}
\author{Hugh~G.~A.~Burton}
\email{hugh.burton@chem.ox.ac.uk}
\affiliation{\UOX}
\date{\today}

\begin{abstract}
Matrix elements between nonorthogonal Slater determinants represent an essential 
component of many emerging electronic structure methods.
\hugh{However, evaluating nonorthogonal matrix elements is conceptually and computationally
harder then their orthogonal counterparts.
While several different approaches have been developed,
these are predominantly derived from the first-quantised generalised Slater--Condon rules 
and usually require biorthogonal occupied orbitals to be computed for each matrix element.
For coupling terms between nonorthogonal excited configurations, a second-quantised
approach such as the nonorthogonal Wick's theorem is more desirable, but this
fails when the two reference determinants have a zero many-body overlap. }
In this contribution, we derive an entirely generalised extension 
to the nonorthogonal Wick's theorem that is applicable to all pairs of determinants with
nonorthogonal orbitals.
\hugh{Our approach creates a universal methodology for evaluating any nonorthogonal matrix element
and allows Wick's theorem and the generalised Slater--Condon rules to be unified for the first time. 
Furthermore, we present a simple well-defined protocol for deriving arbitrary 
coupling terms between nonorthogonal excited configurations.
In the case of overlap and one-body operators, this protocol recovers efficient formulae
with reduced scaling, promising significant computational acceleration for methods that
rely on such terms.}
\end{abstract}
\maketitle

\raggedbottom
\twocolumngrid


\section{Introduction}
\raggedbottom
\hugh{Matrix elements between nonorthogonal Slater determinants are
increasingly common in emerging electronic structure methods.
For example, capturing strong correlation using a linear combination of nonorthogonal Slater determinants is a relatively old idea%
\cite{Fukutome1988,Koch1993,TenNo1997,Ayala1998} 
that has seen a renaissance in the past decade.%
\cite{Thom2009,Sundstrom2014,Mayhall2014,Oosterbaan2018,Jensen2018,Burton2019c,Huynh2019,Nite2019,Kathir2020,Burton2020}.
Similarly, nonorthogonal matrix elements arise in projected Hartree--Fock methods\cite{Scuseria2011,Tsuchimochi2016} 
while the combination of geminal-based nonorthogonal functions is an area of ongoing research.\cite{Dutta2021} 
In each method, the nonorthogonality of different determinants can capture strong 
static correlation effects by breaking and restoring symmetries of the Hamiltonian,\cite{Scuseria2011}
or it can provide a quasi-diabatic representation of dominant electronic configurations.\cite{Thom2009,Jensen2018}
Alternatively, multiple wave functions built from different orbitals arise in orbital-optimised
excited states identified through methods such as $\Delta$SCF,\cite{Gilbert2008,Hait2020,CarterFenk2020,Levi2020}
excited-state mean-field theory,\cite{Shea2018,Hardikar2020} 
or the complete active space self-consistent field.\cite{Tran2019,Tran2020}
In these cases, orbital optimisation can significantly improve predictions of charge transfer excitations, 
but nonorthogonal matrix elements are required for inter-state coupling terms such as oscillator strengths.}

\hugh{A variety of different approaches have been developed for the efficient evaluation of nonorthogonal matrix 
elements,\cite{Leasure1985,Verbeek1991,Igawa1995,Utsuno2013,RodriguezLaguna2020}
 which are predominantly derived from L\"{o}wdin's general formula.\cite{Lowdin1955}
The most popular framework in quantum chemistry is the generalised Slater--Condon rules,\cite{Verbeek1991,MayerBook}
where biorthogonal occupied orbitals are constructed\cite{Amos1961,Hall1951} 
and a modified form of the Slater--Condon rules\cite{SzaboBook} is applied
depending on the number of zero-overlap orbital pairs in the biorthogonal basis.
This approach is applicable to any pair of determinants, but 
requires the diagonalisation of the occupied orbital overlap matrix each time.
In contrast, the development of many-body correlation methods using orthogonal determinants has 
greatly benefited from the second-quantised Wick's theorem.\cite{ShavittBook}
While a nonorthogonal variant of Wick's theorem exists, it is limited to determinants that have 
a non-zero many-body overlap and is not applicable if there are zero-overlap orbital pairs 
in the biorthogonal basis.\cite{Hendekovic1981,RingBook}
This limitation arises because the Thouless theorem,\cite{Thouless1960} used to relate two nonorthogonal determinants 
via an exponential transformation, breaks down when the two determinants have a zero many-body overlap.
As a result, the nonorthogonal Wick's theorem has seen only limited use in 
quantum chemistry.\cite{Jimenez-Hoyos2012a,Tsuchimochi2016,Tsuchimochi2016b}}

Computationally efficient nonorthogonal matrix elements become increasingly 
important in methods that use orthogonally-excited configurations from 
nonorthogonal reference determinants. 
For example, including post-NOCI dynamic correlation in methods such as perturbative 
NOCI-MP2\cite{Yost2013,Yost2016,Yost2019} and NOCI-PT2,\cite{Burton2020} 
or nonorthogonal multireference CI,\cite{TenNo1997,Tsuchimochi2016,Tsuchimochi2016b,Nite2019b}
requires overlap, one-body, or two-body coupling terms between excitations from nonorthogonal determinants.
The number of nonorthogonal matrix elements therefore grows rapidly, and repeated
biorthogonalisation of the occupied orbitals becomes prohibitively expensive.
In principle, the nonorthogonal Wick's theorem could allow these matrix elements
to be evaluated using only biorthogonal reference orbitals
but, until now, this requires the reference determinants to have a strictly non-zero overlap.

In this contribution, we derive an entirely generalised nonorthogonal form of Wick's theorem 
that applies to any pair of determinants with nonorthogonal orbitals, even if the overall 
determinants have a zero overlap.
This new framework, which we call the ``Extended Nonorthogonal Wick's Theorem'', provides
the most general approach for deriving matrix elements using second-quantisation, allowing 
Wick's theorem and the generalised Slater--Condon rules to be unified for the first time.

\hugh{
One particular advantage of our approach is that it allows all matrix elements between excited configurations from a pair 
of nonorthogonal determinants 
to be derived using a single well-defined protocol.
While some of the resulting expressions have previously been derived for various bespoke 
applications (see e.g.\ Refs.~\onlinecite{Yost2013,Sundstrom2014,Nite2019b}),
 these have often relied on
the properties of matrix determinants to account for orbital excitations. 
Instead, we present a unifying theory that can recover all of these results and is automatically applicable
in cases where the reference determinants have a zero overlap.
Furthermore, we show how evaluating intermediates for a given pair of determinants
can reduce the scaling of overlap and one-body coupling terms between excited configurations to $\Or{1}$.
These particular non-orthogonal matrix elements then become almost as straightforward
as the orthogonal Slater--Condon rules or Wick's theorem, promising
considerable acceleration for methods that rely on such terms.}

To derive our generalised nonorthogonal matrix elements, we first define our notation in 
Section~\ref{sec:Notation}.
In Section~\ref{sec:ExThoulessTrans}, we extend Thouless' theorem\cite{Thouless1960} 
to the case where the two determinants have nonorthogonal orbitals and a zero many-body overlap.
Section~\ref{sec:ExWickTheorem} combines this extended Thouless transformation with Wick's theorem 
to create a generalised protocol for evaluating nonorthogonal matrix elements using second-quantisation.
We then illustrate the application of this approach by re-deriving the generalised Slater--Condon rules
for one- and two-body operators in Section~\ref{sec:gsc}.
Finally, in Section~\ref{sec:ExcitedConfig}, we extend our framework to the matrix elements between excited configurations
and show how $\Or{1}$ scaling can be achieved for overlap and one-body operators.

\section{Notation}
\label{sec:Notation}
We will consider matrix elements between the two determinants $\ket{^{w}\Phi}$ and $\ket{^{x}\Phi}$.
Each determinant is constructed from a bespoke set of molecular orbitals (MOs), 
represented in terms of the atomic spin-orbital basis functions $\ket{\chi_{\mu}}$ as
\begin{align}
\ket{^{w}\phi_p} = \sum_{\mu} \ket{\chi_{\mu}}\, ^{w}\C^{\mu \cdot}_{\cdot p}.
\label{eq:spinOrbital}
\end{align}
Here, we employ the nonorthogonal tensor notation of Head-Gordon \etal{}\cite{HeadGordon1998} 
\hugh{to explicitly keep track of any required overlap matrices.}
\hugh{Occupied MOs are indexed} as ($i,j,k,$ etc), virtual MOs as ($a,b,c,$ etc), 
and any general MO as ($p,q,r,$ etc).
We emphasise that the MOs are orthogonal \emph{within} each Slater determinant, but are 
nonorthogonal \emph{between} the different determinants.

In second-quantisation, the $\Ne$-electron determinant $\ket{^{w}\Phi}$ is defined as
\begin{align}
\ket{^{w}\Phi} = \prod_{i=1}^{\Ne} \, ^{w}\bdag_{i} \vac,
\end{align}
where $\ket{-}$ is the physical vacuum and the molecular orbital creation operators 
$^{w}\bdag_{i}$ satisfy the standard fermionic anticommutation rules.\cite{HelgakerBook}
Using the expansion Eq.~\eqref{eq:spinOrbital}, the MO creation and annihilation operators 
can be represented in terms of the \hugh{covariant} atomic spin-orbital creation $\adag_{\mu}$ and 
annihilation $\a_{\mu}$ operators as
\begin{align}
	^{w}\bdag_{p} = \sum_{\mu} \adag_{\mu}\, ^{w}\C^{\mu \cdot}_{\cdot p} \quad\text{and}\quad 
	^{w}\b_{p}      = \sum_{\mu} (^{w}\C^{*})^{\cdot \mu}_{p \cdot} \a_{\mu}.
\label{eq:MObOperators}
\end{align}
The covariant atomic spin-orbital operators have only one non-zero anticommutator\cite{HelgakerBook}
\begin{align}
	\acomm{\adag_{\nu}}{\a_{\mu}} = g_{\mu \nu},
\end{align}
where $g_{\mu \nu} = \brkt{\chi_{\mu}}{\chi_{\nu}}$ defines the corresponding 
covariant metric tensor (overlap matrix).\cite{HeadGordon1998}

\hugh{Throughout this paper, we will need to express the atomic spin-orbital creation and annihilation 
operators in terms of the MO creation and annihilation operators using the inverse of Eq.~\eqref{eq:MObOperators}
\begin{subequations}
\begin{align}
    \adag_{\mu} &= \sum_{p\sigma} \up{w}\bdag_{p}\, (\up{w}\C^{*})^{\cdot \sigma}_{p \cdot} g_{\sigma \mu} 
    \\
    \a_{\mu}    &= \sum_{p\sigma} g_{\mu \sigma}(\up{w}\C^{\vphantom{*}})^{\sigma \cdot}_{\cdot p}\, \up{w}\b_{p}.
\end{align}
\label{eq:AOaOperators}
\end{subequations}
To avoid the introduction of overlap matrices throughout our expressions, we will often use the 
contravariant atomic spin-orbital operators $(a^{\mu})^{\dagger}$ and $a^{\mu}$ defined as
\begin{equation}
(a^{\nu})^\dagger = \sum_{\mu} a_{\mu}^{\dagger}\, g^{\mu \nu}  \quad\text{and}\quad
a^{\nu}= \sum_{\mu} g^{\nu \mu} a_{\mu}, 
\label{eq:AOcontra}
\end{equation}
}
where, $g^{\mu \nu}$ is the contravariant metric tensor corresponding to the inverse
covariant overlap matrix,\cite{HeadGordon1998} \ie{} 
\begin{align}
    g^{\mu \nu} = (g^{-1})_{\mu \nu}.
\end{align}
\hugh{If the AO basis is overcomplete, this contravariant metric tensor becomes the pseudo-inverse
of the covariant overlap matrix.}
Note that the anticommutator of these contravariant atomic spin-orbital operators is
\begin{align}
	\acomm{(a^{\nu})^{\dagger}}{a^{\mu}} = g^{\mu \nu}.
	\label{eq:covariantAnticomm}
\end{align}

\section{Extended Thouless Transformation}
\label{sec:ExThoulessTrans}

\subsection{Conventional Thouless Transformation}

The conventional form of Thouless' theorem allows two nonorthogonal determinants to be
related by an exponential operator of single excitations as\cite{Thouless1960}
\begin{align}
\ket{^{w}\Phi} = \exp(\cZ) \ket{^{x}\Phi} \brkt{^{x}\Phi}{^{w}\Phi}.
\label{eq:ThoulessTheorem}
\end{align} 
To derive the single excitation operator $\cZ$, the occupied orbitals can be
be transformed to a biorthogonal basis using L\"owdin pairing\cite{Amos1961,Hall1951} such that 
\begin{align}
\sum_{\mu \nu} (^{x}\tC^{*})^{\cdot \mu}_{i \cdot}\, g_{\mu \nu} \up{w}\tC^{\nu \cdot}_{\cdot j} = \hugh{\up{xw}\tilde{S}_{i}} \delta_{ij}.
\label{eq:LowdinPairing}
\end{align}
\hugh{The virtual-occupied and virtual-virtual blocks of the biorthogonalised overlap matrix become
\begin{align}
\sum_{\mu \nu} (^{x}\tC^{*})^{\cdot \mu}_{a \cdot}\, g_{\mu \nu} \up{w}\tC^{\nu \cdot}_{\cdot p} = \up{xw}\tilde{S}_{ap},
\label{eq:PairedOccVir}
\end{align}}
and the transformed molecular orbital creation and annihilation operators are given as 
\begin{align}
	^{w}\tbdag_{p} = \sum_{\mu} \adag_{\mu}\, ^{w}\tC^{\mu \cdot}_{\cdot p} \quad\text{and}\quad 
	^{w}\tb_{p}      = \sum_{\mu} (^{w}\tC^{*})^{\cdot \mu}_{p \cdot}\, \a_{\mu}.
\label{eq:tMObOperators}
\end{align}
The single excitation operator \hugh{in Eq.~\eqref{eq:ThoulessTheorem}} is then defined as
\begin{align}
\cZ = \sum_{ia} \up{xw}Z_{ai} \, ^{x}\tbdag_{a}  \, ^{x}\tb_{i}.
\end{align}
with the $^{xw}Z_{ai}$ matrix elements given by 
\begin{align}
^{xw}Z_{ai} = \sum_{\mu \nu} (^{x}\tC^{*})^{\cdot \mu}_{a \cdot}\, g_{\mu \nu}\, (^{w}\tC)^{\nu \cdot}_{\cdot i} \frac{1}{\hugh{\up{xw}\tilde{S}_{i}}}.
\label{eq:Zdef}
\end{align}
A brief derivation of this result can be found in Appendix~\ref{apdx:ThoulessThereom}.

Unfortunately, this exponential representation relies on the strict nonorthogonality of the two determinants 
$\brkt{^{x}\Phi}{^{w}\Phi} \neq 0$; in other words, it is not applicable to a pair of determinants
that are orthogonal but contain mutually nonorthogonal orbitals.
Our first step is therefore a generalisation of the Thouless transformation to the case
where $\brkt{^{x}\Phi}{^{w}\Phi} = 0$.

\subsection{Introducing Zero-Overlap Orbitals}
We begin in the biorthogonal basis identified through L\"owdin pairing, with orbital 
coefficients satisfying Eq.~\eqref{eq:LowdinPairing}.
For a general pair of nonorthogonal orbitals, it is possible for orbital
pairs to have a zero-overlap in the biorthogonal basis, where \hugh{$\up{xw}\tilde{S}_{i} = 0$}.
Taking the case with $m$ zero overlaps between orbitals $k_1, \cdots, k_m$, 
we construct ``reduced'' determinants by removing the electrons in these zero-overlap
orbitals to give
\begin{subequations}
\begin{align}
\ket{^{x}\Phi_{k_{1} \cdots k_{m}}} &= \up{x}\tb_{k_{m}}\cdots\up{x}\tb_{k_{1}}  \ket{^{x}\Phi}
\\
\ket{^{w}\Phi_{k_{1} \cdots k_{m}}} &= \up{w}\tb_{k_{m}}\cdots\up{w}\tb_{k_{1}}  \ket{^{w}\Phi}.
\end{align}
\end{subequations}
These reduced determinants are strictly nonorthogonal with the non-zero reduced overlap defined as
\begin{align}
\up{xw}\tilde{S} = \brkt{^{x}\Phi_{k_{1} \cdots k_{m}}}{^{w}\Phi_{k_{1} \cdots k_{m}}} = \prod_{ \{i |  \up{xw}\tilde{S}_{i} \neq 0\} } \hugh{\up{xw}\tilde{S}_{i}}.
\end{align}
Therefore, the Thouless transformation can now be applied to these reduced determinants to give
\begin{align}
	\ket{^{w}\Phi_{k_{1} \cdots k_{m}}} = \exp (\tilde{\cZ}) \ket{^{x}\Phi_{k_{1} \cdots k_{m}}} \up{xw}\tilde{S}.
\end{align}
Here, we have introduced the reduced single excitation operator $\tilde{\cZ}$ that only contains 
excitations from occupied orbitals with a non-zero overlap as
\begin{align}
\tilde{\cZ} = \sum_{ \{i |  \up{xw}\tilde{S}_{i} \neq 0\} } \sum_{a} \up{xw}Z_{ai} \, ^{x}\tbdag_{a}  \, ^{x}\tb_{i} .
\label{eq:reducedSingleEx}
\end{align}
The full $\Ne$-electron determinants are then related through second-quantisation as
\begin{align}
\ket{^{w}\Phi} = \up{w}\tbdag_{k_{1}}\cdots\up{w}\tbdag_{k_{m}} \exp (\tilde{\cZ})  \up{x}\tb_{k_{m}}\cdots\up{x}\tb_{k_{1}} \ket{^{x}\Phi} \up{xw}\tilde{S}.
\label{eq:LongForm}
\end{align}

Equation~\eqref{eq:LongForm} can be further simplified by exploiting the commutativity relation
$[\tilde{\cZ},\up{x}\tb_{k}]=0$ to shift the $ \exp (\tilde{\cZ})$ operator to the far right-hand side, giving
\begin{align}
\ket{^{w}\Phi} = \up{w}\tbdag_{k_{1}}\cdots\up{w}\tbdag_{k_{m}}  \up{x}\tb_{k_{m}}\cdots\up{x}\tb_{k_{1}} \exp (\tilde{\cZ})  \ket{^{x}\Phi} \up{xw}\tilde{S}.
\label{eq:ShiftedLongForm}
\end{align}
We can then introduce single-electron excitation operators $\hat{z}_{k}$ for the zero-overlap orbitals as
\begin{align}
\begin{split}
	\hat{z}_{k} = \hspace{0pt}^{w}\tbdag_{k}{^{x}} \tb_{k} 
	&= \sum_{p} \Big(\sum_{\mu \nu} (^{x}\tC^{*})^{\cdot \mu}_{p \cdot} g_{\mu \nu} (^{w}\tC)^{\nu \cdot}_{\cdot k} \Big) \up{x}\tbdag_{p} \up{x}\tb_{k}
	\\
	&= \sum_{a} \hugh{\up{xw}\tilde{S}_{a k}}\, \up{x}\tbdag_{a} \up{x}\tb_{k},
\end{split}
\end{align}
where we have exploited the biorthogonality and zero-overlap
of the occupied orbitals such that \hugh{$\up{xw}\tilde{S}_{i k} = 0$} for all $i$. 
The commutativity of these single excitation operators with the $^{x}\tb_{k}$ annihilation operators
leads to the simplified relationship 
\begin{align}
\ket{^{w}\Phi} =  \prod_{ \{k |  \up{xw}\tilde{S}_k = 0\} }  \hat{z}_{k}  \exp (\tilde{\cZ}) \ket{^{x}\Phi} \up{xw}\tilde{S}.
\label{eq:ProdForm}
\end{align}

Introducing the relationship $\hat{z}_{k} = \exp(\hat{z}_{k}) - 1$ allows Eq.~\eqref{eq:ProdForm} to
be expanded as
\begin{align}
\ket{^{w}\Phi} = \prod_{ \{k |  \up{xw}\tilde{S}_k = 0\} } \Big( \exp(\hat{z}_{k}) - 1 \Big) \exp \big(\tilde{\cZ} \big) \ket{^{x}\Phi} \up{xw}\tilde{S}.
\end{align}
Expanding the product of $(\exp(\hat{z}_k) - 1)$ terms then leads to a sum of exponential transformations
where every combination of the zero-overlap single excitation operators $\hat{z}_k$ is either included or excluded 
with an appropriate phase factor, giving
\begin{align}
\ket{^{w}\Phi}  = \sum_{n=0}^{m} (-1)^{(m-n)} \sum_{\Cn} \exp \big(\tilde{\cZ}^{\Cn} \big) \ket{^{x}\Phi} \up{xw}\tilde{S}.
\label{eq:ExtendedThouless}
\end{align}
Here, we have introduced the compound index $\cC_n$ to denote a particular combination of $n$ out of $m$ zero-overlap orbitals, 
while the superscript notation $\tilde{\cZ}^{\cC_n}$ indicates which particular zero-overlap excitations 
are included in the corresponding operator, \ie{}
\begin{align}
\tilde{\cZ}^{\Cn} &= \tilde{\cZ} + \sum_{k \in \Cn} \hat{z}_{k}.
\label{eq:zero_transformed_excitation}
\end{align}
We refer to this transformation in its various forms \eqref{eq:ProdForm}--\eqref{eq:ExtendedThouless}
as the {``Extended Thouless Transformation''}.
To explicitly illustrate its application, the case of two zero-overlaps in orbital pairs $k_1$ and $k_2$ leads to 
\begin{align}
\begin{split}
\ket{^{w}\Phi} &= \up{xw}\tilde{S} \times
\\
\Big(&\exp \tilde{\cZ}^{(k_1, k_2)}  - \exp \tilde{\cZ}^{(k_1)}  - \exp  \tilde{\cZ}^{(k_2)}
+\exp\tilde{\cZ} \Big) 	\ket{^{x}\Phi},
\end{split}
\label{eq:ExtendedThoulessExample}
\end{align}
where $\tilde{\cZ}^{(k_1, k_2)} = \tilde{\cZ} + \hat{z}_{k_1} + \hat{z}_{k_2}$ and $\tilde{\cZ}^{(k_1)} = \tilde{\cZ}  + \hat{z}_{k_1}$.
\\

\section{Extended Nonorthogonal Wick's Theorem}
\label{sec:ExWickTheorem}
\subsection{Conventional Wick's Theorem}
Efficiently deriving matrix elements using the conventional Wick's theorem requires the introduction of contractions, defined 
for two creation or annihilation operators $\up{x}b_{p}$ and $\up{x}b_{q}$ as
\begin{align}
    \wick{ \c1{\up{x}b_p} \c1{\up{x}b_{q}} } = \up{x}b_p \up{x}b_q - \{ \up{x}b_p \up{x}b_q \},
\end{align}
where $\{ \up{x}b_p \up{x}b_q \}$ represents a normal-ordered operator string with respect to the
reference Fermi vacuum $\braket{\up{x}\Phi}{\cdots}{\up{x}\Phi}$.\cite{ShavittBook}
The only non-zero contractions between creation and annihilation operators with respect
to this symmetric Fermi vacuum are
\begin{align}
\wick{\up{x} \c1 b_{a}\up{x} \c1 b^{\dagger}_{b}} = \delta_{ab}
\quad \text{and} \quad
\wick{\up{x} \c1 b^{\dagger}_{i}\up{x} \c1 b_{j}} = \delta_{ij}.
\label{eq:PureContractions}
\end{align}
Through Wick's theorem, the Fermi vacuum expectation of an operator product is
given by the sum over all fully contracted products of operators, e.g.\
\begin{align}
\begin{split}
\brkt{\up{x}\Phi_{i}^{a}}{\up{x}\Phi_{j}^{b}}
=
\wick{\braket{\up{x}\Phi}{\up{x}\c2 b_i^{\dagger} \up{x}\c1 b_a \up{x}\c1 b_b^{\dagger} \up{x}\c2 b_j}{\up{x}\Phi}}
=\delta_{ij} \delta_{ab}.
\end{split}
\end{align}

\subsection{Zero-Overlap Transformed Operators}
Using the extended Thouless transformation, we can now extend the nonorthogonal Wick's theorem\cite{RingBook,Balian1969,Hendekovic1981}
to derive matrix elements between \emph{any} pair of determinants with mutually nonorthogonal orbitals.
\hugh{In what follows, we will consider the contravariant atomic spin-orbital operators (see Section~\ref{sec:Notation})
to avoid large numbers of overlap matrices in our expressions.}
The matrix elements for general operators expressed in the atomic spin-orbital basis
requires the evaluation of terms containing a string of creation and annihilation operators, such as
$\braket{^{x}\Phi}{(a^{\mu})^{\dagger} (a^{\nu})^{\dagger} \cdots a^{\sigma} a^{\tau}}{^{w}\Phi}$.
Applying the extended Thouless transformation leads to the linear combination
\begin{align}
&\braket{^{x}\Phi}{(a^{\mu})^{\dagger} (a^{\nu})^{\dagger} \cdots a^{\sigma} a^{\tau}}{^{w}\Phi} \label{eq:FullMatrixElement}
=
 \up{xw}\tilde{S} \times
\\
&\sum_{n=0}^{m} (-1)^{(m-n)} 
\sum_{\cC_n}\braket{^{x}\Phi}{(a^{\mu})^{\dagger} (a^{\nu})^{\dagger} \cdots a^{\sigma} a^{\tau} \exp \big( \tilde{\cZ}^{\cC_n} \big) }{^{x}\Phi}.
\nonumber
\end{align}
To evaluate each constituent matrix element for the combinations $\Cn$, we follow the approach described in 
Refs.~\onlinecite{RingBook} and \onlinecite{Jimenez-Hoyos2012a}
and introduce a similarity-transformed set of spin-orbital creation 
and annihilation operators as
\begin{subequations}
\begin{align}
(d[\mathsmaller{\Cn}]^{\mu})^{\dagger}  &= \exp \big(-\tilde{\cZ}^{\cC_n} \big) (a^{\mu})^{\dagger} \exp \big(\tilde{\cZ}^{\cC_n} \big)
\\
d[\mathsmaller{\Cn}]^{\mu} &= \exp \big(-\tilde{\cZ}^{\cC_n} \big) a^{\mu} \exp \big(\tilde{\cZ}^{\cC_n} \big).
\end{align}
\label{eq:TransformedOperators}
\end{subequations}
These operators clearly depend on the particular combination $\Cn$ 
of included zero-overlap single excitation operators.
Expanding the similarity transformation as
\begin{align}
\exp\big(-\tilde{\cZ}^{\cC_n} \big) \hugh{a^{\mu}} \exp \big(\tilde{\cZ}^{\cC_n} \big) 
&= \hugh{a^{\mu}} - [\tilde{\cZ}^{\cC_n}, \hugh{a^{\mu}}], 
\end{align}
and similarly for $(\hugh{a^{\mu}})^{\dagger}$, leads to the explicit forms
\begin{widetext}
\begin{subequations}
\begin{align}
(d[\mathsmaller{\Cn}]^{\mu})^{\dagger}  
&= \sum_i \up{x} \tbdag_{i} (^{x}\tC^{*})^{\cdot \mu}_{i \cdot}
+ \sum_a \up{x}\tbdag_{a} \Big[ (^{x}\tC^{*})^{\cdot \mu}_{a \cdot} 
- \sum_{ \{i | \up{xw}\tilde{S}_i \neq 0 \} }\hspace{-5pt} \up{xw}Z_{ai} (^{x} \tC^{*})^{\cdot \mu}_{i \cdot} 
- \sum_{k \in \cC_n} \hugh{\up{xw}\tilde{S}_{ak}} (^{x} \tC^{*})^{\cdot \mu}_{k \cdot}\Big],
\label{eq:TransformedOperators2a}
\\
d[\mathsmaller{\Cn}]^{\mu} 
&= \sum_i  (^{x}\tC)^{\mu \cdot}_{\cdot i}\up{x}\,\tb_{i}  
+ \sum_a \Big[ \sum_{\{i | \up{xw}\tilde{S}_{i} \neq 0 \}} \hspace{-5pt} (^{x}\tC)^{\mu \cdot}_{\cdot a}\, \up{xw}Z_{ai} \up{x}\tb_{i} 
+ \sum_{k \in \cC_n}  (^{x}\tC)^{\mu \cdot}_{\cdot a}\, \hugh{\up{xw}\tilde{S}_{ak}}\, \hugh{\up{x}\tb_{k}} 
+  (^{x}\tC)^{\mu \cdot}_{\cdot a}\,  \hugh{\up{x}\tb_{a}} \Big].
\label{eq:TransformedOperators2b}
\end{align}
\label{eq:TransformedOperatorsExpand}
\end{subequations}
\end{widetext}
An explicit derivation of these relationships can be found in Appendix~\ref{apdx:SimTransform}.
Exploiting the relationship
\begin{align}
	\bra{^{x}\Phi} \exp \big( -\tilde{\cZ}^{\cC_n} \big) = \bra{^{x}\Phi} 
\end{align}
and \hugh{the resolution of the identity} 
\begin{align}
	\exp \big(\tilde{\cZ}^{\cC_n} \big) \exp \big(-\tilde{\cZ}^{\cC_n} \big) = \cI
\end{align}
then allows the constituent matrix elements within Eq.~\eqref{eq:FullMatrixElement} to be expressed as
\begin{align}
\begin{split}
\braket{^{x}\Phi}{(a^{\mu})^{\dagger} (a^{\nu})^{\dagger} &\cdots a^{\sigma} a^{\tau} \exp \big(\tilde{\cZ}^{\cC_n} \big)}{^{x}\Phi}
=
\\ &\braket{^{x}\Phi}{(d[\mathsmaller{\Cn}]^{\mu})^{\dagger} (d[\mathsmaller{\Cn}]^{\nu})^{\dagger} \cdots d[\mathsmaller{\Cn}]^{\sigma} d[\mathsmaller{\Cn}]^{\tau}}{^{x}\Phi}.
\end{split}
\label{eq:TransformedMatrixElement}
\end{align}


\subsection{The Fundamental Contraction}
\label{subsec:FundamentalContraction}
The extended Thouless transformation essentially converts the nonorthogonal matrix element with an 
asymmetric Fermi vacuum $\braket{\up{x}\Phi}{\cdots}{\up{w}\Phi}$ to a transformed matrix element with respect
to symmetric Fermi vacuum $\braket{\up{x}\Phi}{\cdots}{\up{x}\Phi}$.
Since the transformed operators $d[\Cn]^{\mu}$ and $(d[\Cn]^{\mu})^{\dagger}$ are expressed purely in terms of 
the $\up{x}\tb_p$ creation and $\up{x}\tbdag_p$ annihilation operators,
with respect to the $\braket{\up{x}\Phi}{\cdots}{\up{x}\Phi}$ vacuum, 
their non-zero contractions with respect to $\braket{\up{x}\Phi}{\cdots}{\up{x}\Phi}$ can be derived by  
combining Eqs.~\eqref{eq:PureContractions} and \eqref{eq:TransformedOperatorsExpand} to give 
\begin{subequations}
\begin{align}
\wick{(\c1 d[\mathsmaller{\Cn}]^{\mu})^{\dagger} (\c1 d[\mathsmaller{\Cn}]^{\nu})}
&= \up{xw}M^{\nu \mu} - \sum_{k \notin \cC_n}  \up{xw}P_{k}^{\nu \mu},
\label{eq:LeftContraction}
\\
\wick{(\c1 d[\mathsmaller{\Cn}]^{\mu}) (\c1 d[\mathsmaller{\Cn}]^{\nu})^{\dagger}}
&= g^{\mu \nu} - \up{xw}M^{\mu \nu} + \sum_{k \notin \cC_n}  \up{xw}P_{k}^{\mu \nu},
\label{eq:RightContraction}
\end{align}
\end{subequations}
where we have introduced the general notation
\begin{subequations}
\begin{align}
\up{xw}P_{k}^{\nu \mu} &= (^{w}\tC)^{\nu \cdot}_{\cdot k} (^{x}\tC^{*})^{\cdot \mu}_{k \cdot},
\label{eq:CoDensity}
\\
\up{xw}P^{\nu \mu} &= \sum_{\{k| \up{xw}\tilde{S}_{k} = 0\} } \hspace{-2pt}^{xw}P_{k}^{\nu \mu},
\label{eq:CoDensitySum}
\\
\up{xw}W^{\nu \mu} &= \sum_{\{ i | \up{xw}\tilde{S}_{i} \neq 0\} } (^{w}\tC)^{\nu \cdot}_{\cdot i} \frac{1}{\up{xw}\tilde{S}_{i}} (^{x}\tC^{*})^{\cdot \mu}_{i \cdot},
\\
\up{xw}M^{\nu \mu} &=  \up{xx}P^{\nu \mu} + \up{xw}P^{\nu \mu} + \up{xw}W^{\nu \mu}.
\label{eq:Msum}
\end{align}
\end{subequations}
\hugh{These expressions closely resemble co-density matrices,\cite{Thom2009} and their derivation
can be found in Appendix~\ref{apdx:SimTransformContract}.}

The overall matrix element $\braket{^{x}\Phi}{(a^{\mu})^{\dagger} (a^{\nu})^{\dagger} \cdots a^{\sigma} a^{\tau}}{^{w}\Phi}$
 requires the derivation of contractions between 
the $a^{\mu}$ and $(a^{\mu})^{\dagger}$ \hugh{atomic spin-orbital} operators rather than the transformed $d[\Cn]^{\mu}$ and
$(d[\Cn]^{\mu})^{\dagger}$ operators.  
To show how a general string of operators can be evaluated using \hugh{the contractions defined 
Eqs~\eqref{eq:LeftContraction} and \eqref{eq:RightContraction}}, we first demonstrate
the derivation of the one-body co-density matrix element
\begin{align}
\up{xw}\Gamma_1^{\nu \mu} = \braket{^{x}\Phi}{(a^{\mu})^{\dagger} a^{\nu}}{^{w}\Phi}.
\end{align}
Assuming that some form of Wick's theorem can be derived, we expect this matrix element to be represented by
the single contraction
\begin{align}
\up{xw}\Gamma_1^{\nu \mu} = \braket{^{x}\Phi}{\wick{(\c1 a^{\mu})^{\dagger} \c1 a^{\nu}}}{^{w}\Phi}.
\end{align}
Taking the most general case with $m$ zero-overlap orbitals,
the extended Thouless transformation leads to 
\begin{widetext}
\begin{align}
    \begin{split}
\braket{^{x}\Phi}{\wick{(\c1 a^{\mu})^{\dagger} \c1 a^{\nu}}}{^{w}\Phi} 
&=
\up{xw}\tilde{S}\sum_{n=0}^{m} (-1)^{(m-n)}
\sum_{\cC_n} 
\braket{^{x}\Phi}{%
    \wick{(\c1 d[\mathsmaller{\Cn}]^{\mu})^{\dagger} \c1 d[\mathsmaller{\Cn}]^{\nu}}%
}{^{x}\Phi}
\\&
=
\up{xw}\tilde{S}\sum_{n=0}^{m} (-1)^{(m-n)}
\sum_{\cC_n} \qty( \up{xw}M^{\nu \mu} - \sum_{k \notin \Cn} \up{xw}P_{k}^{\nu \mu} ).
\end{split}
\label{eq:oneElectronExpansion}
\end{align}
Reversing the order of summation over $k$ and $\Cn$ yields
\begin{align}
\braket{^{x}\Phi}{\wick{(\c1 a^{\mu})^{\dagger} \c1 a^{\nu}}}{^{w}\Phi} 
=
\up{xw}\tilde{S}\sum_{n=0}^{m} (-1)^{(m-n)} \qty(
\ \up{xw}M^{\nu \mu} \bigg[\sum_{\Cn}\ 1 \bigg] 
- \sum_{k} \up{xw}P_{k}^{\nu \mu} \bigg[\sum_{\Cn \niton\, k}\ 1 \bigg] ).
\label{eq:oneElectronSqBrackets}
\end{align}
The first term in square brackets $\qty[\sum_{\Cn}1]$ can be recognised as the total number of ways to pick $n$ orbitals 
from the $m$ zero-overlap orbitals, given by ${m \choose n}$.
Similarly, the second term in square brackets $\qty[\sum_{\Cn \niton\, k}1]$ is the total number of ways to pick $n$ orbitals 
from the $m-1$ zero-overlap orbitals that remain when orbital $k$ is excluded, given by ${m-1 \choose n}$.
These combinatorial expansions allow Eq.~\eqref{eq:oneElectronSqBrackets} to be expressed as
\begin{align}
\braket{^{x}\Phi}{\wick{(\c1 a^{\mu})^{\dagger} \c1 a^{\nu}}}{^{w}\Phi} 
=
\up{xw}\tilde{S} \qty[\sum_{n=0}^{m} (-1)^{(m-n)}
 \binom{m}{n} \up{xw}M^{\nu \mu}
+ \sum_{n=0}^{m-1} (-1)^{(m-1-n)} \binom{m-1}{n} \up{xw}P^{\nu \mu} ].
\label{eq:oneElectronBinomialRearrange}
\end{align}
\end{widetext}
Here, we note that there are no ways to exclude an orbital $k$ when all zero-orbital overlaps are included
in the complete combination $\cC_m$.
Exploiting the binomial expansion
\begin{align}
    \sum_{n=0}^{y} (-1)^{(y-n)} {y \choose n} = \hugh{\delta_{0,y}} 
    \label{eq:binomial}
\end{align}
then leads to the closed-form expression
\begin{align}
\begin{split}
    \braket{^{x}\Phi}{&\wick{(\c1 a^{\mu})^{\dagger} \c1 a^{\nu}}}{^{w}\Phi} = 
\up{xw}\tilde{S} \Big[ \hugh{\delta_{0,m}}\ \up{xw}M^{\nu \mu} + \hugh{\delta_{0,m-1}}\, ^{xw}P^{\nu \mu} \Big].
\end{split}
\label{eq:ExpandedOneDensity}
\end{align}
The reduced overlap $\up{xw}\tilde{S}$ will be a prefactor for every matrix element between these 
nonorthogonal determinants. 
The remaining terms can then be used to define ``{fundamental contractions}'' for second-quantisation
operators with respect to the asymmetric Fermi vacuum $\braket{\up{x}\Phi}{\cdots}{\up{w}\Phi}$.
The form of these contractions depends on the number of zero-overlap orbitals $m$ and
 we define the first fundamental contraction as
\begin{align}
    \wick{(\c1 a^{\mu})^{\dagger} \c1 a^{\nu}} =
	\begin{cases}
	\up{xw}M^{\nu \mu} 	  & m=0,
	\\
	\up{xw}P^{\nu \mu}    & m=1,
	\\
	0					  & m >1.
	\end{cases}
\label{eq:fun_contract}
\end{align}
Similarly, the second fundamental contraction can be identified as
\begin{align}
    \wick{\c1 a^{\mu} (\c1 a^{\nu})^{\dagger}} =
	\begin{cases}
	g^{\mu \nu} - \up{xw}M^{\mu \nu} 	  & m=0,
	\\
	-\, \up{xw}P^{\mu \nu}    & m=1,
	\\
	0					  & m >1.
	\end{cases}
\label{eq:fun_contract2}
\end{align}
Crucially, we emphasise that these fundamental contractions are defined with respect to the asymmetric
Fermi vacuum $\braket{\up{x}\Phi}{\cdots}{\up{w}\Phi}$.

\subsection{Combining Several Contractions}
Next, we show how the fundamental contractions can be combined to derive matrix elements 
for longer products of creation and annihilation operators.
As an example, consider the two-body reduced co-density matrix element, defined as
\begin{align}
\up{xw}\Gamma_2^{\tau \mu \sigma \nu } = \braket{\up{x}\Phi}{(a^{\mu})^{\dagger}(a^{\nu})^{\dagger}a^{\sigma}a^{\tau}}{\up{w}\Phi}.
\end{align}
Applying Wick's theorem, this matrix element should be given by the sum of the two contractions
\begin{align}
    \braket{\up{x}\Phi}{\wick{(\c2 a^{\mu})^{\dagger}(\c1 a^{\nu})^{\dagger} \c1 a^{\sigma} \c2 a^{\tau}}}{\up{w}\Phi}
    + \braket{\up{x}\Phi}{\wick{(\c1 a^{\mu})^{\dagger}(\c2 a^{\nu})^{\dagger} \c1 a^{\sigma} \c2 a^{\tau}}}{\up{w}\Phi}.
\end{align}
Note that the second term in this expression carries a phase of $-1$ from the intersection of 
the contraction lines, representing the fundamental parity of fermionic operators.\cite{ShavittBook}
Taking the first contraction as an example, we apply the extended Thouless transformation and 
the transformed contractions defined in Eqs.~\eqref{eq:LeftContraction} and \eqref{eq:RightContraction} to give
\begin{widetext}
\begin{align}
\begin{split}
    \braket{\up{x}\Phi}{\wick{(\c2 a^{\mu})^{\dagger}(\c1 a^{\nu})^{\dagger} \c1 a^{\sigma} \c2 a^{\tau}}}{\up{w}\Phi}
    &=
    \up{xw}\tilde{S} \sum_{n=0}^{m} (-1)^{(m-n)} \sum_{\Cn} 
    \braket{\up{x}\Phi}{
    \wick{(\c2 d[\mathsmaller{\Cn}]^{\mu})^{\dagger} (\c1 d[\mathsmaller{\Cn}]^{\nu})^{\dagger} 
           \c1 d[\mathsmaller{\Cn}]^{\sigma} \c2 d[\mathsmaller{\Cn}]^{\tau}}}{\up{x}\Phi}.
   \\
   &=
    \up{xw}\tilde{S} \sum_{n=0}^{m} (-1)^{(m-n)} \sum_{\Cn} 
    \Big(\up{xw}M^{\tau \mu} - \sum_{k_1 \notin \Cn}\up{xw}P_{k_1}^{\tau \mu} \Big)
    \Big(\up{xw}M^{\sigma \nu} - \sum_{k_2 \notin \Cn}\up{xw}P_{k_2}^{\sigma \nu} \Big).
\end{split}
\end{align}
Once again, the reduced overlap $\up{xw}\tilde{S}$ appears as an overall prefactor.
The order of summation over the $k_1, k_2$  indices and $\Cn$ can then be swapped to give
\begin{align}
\begin{split}
    \braket{\up{x}\Phi}{&\wick{(\c2 a^{\mu})^{\dagger}(\c1 a^{\nu})^{\dagger} \c1 a^{\sigma} \c2 a^{\tau}}}{\up{w}\Phi}
    = \up{xw}\tilde{S} \sum_{n=0}^{m} (-1)^{(m-n)} \times
    \\
    &\quad\Big(
    \up{xw}M^{\tau \mu}\ \up{xw}M^{\sigma \nu} \Big[\sum_{\Cn}1 \Big]
    - \sum_{k} (\up{xw}M^{\tau \mu} \ \up{xw}P_{k}^{\sigma \nu} + \ \up{xw}P_{k}^{\tau \mu} \ \up{xw}M^{\sigma \nu} ) 
    \Big[\sum_{\Cn \niton k}1 \Big]
    + \sum_{k_1,k_2} \up{xw}P_{k_1}^{\tau \mu}\ \up{xw}P_{k_{2}}^{\sigma \nu} 
    \Big[\sum_{\Cn \niton k_1,k_2}1 \Big] \Big).
\end{split}
\end{align}
The third term in square brackets $\qty[\sum_{\Cn \niton k_1,k_2}1 ]$ is simply the number of ways to pick $n$ orbitals from the $m-2$ zero-overlap orbitals
that remain when orbitals $k_1$ and $k_2$ are removed, given by ${m-2 \choose n}$.
Applying the binomial expansion Eq.~\eqref{eq:binomial} in an analogous way to the single contraction 
leads to the closed form
\begin{align}
    \braket{\up{x}\Phi}{\wick{(\c2 a^{\mu})^{\dagger}(\c1 a^{\nu})^{\dagger} \c1 a^{\sigma} \c2 a^{\tau}}&}{\up{w}\Phi}
\label{eq:two-body}
    =
    \up{xw}\tilde{S} \times 
    \\
    &\Big[
    \hugh{\delta_{0,m}}
    (\up{xw}M^{\tau \mu}\ \up{xw}M^{\sigma \nu})
    + 
    \hugh{\delta_{0,m-1}}
    (\up{xw}M^{\tau \mu} \ \up{xw}P^{\sigma \nu} + \ \up{xw}P^{\tau \mu} \ \up{xw}M^{\sigma \nu} ) 
    + 
    \hugh{\delta_{0,m-2}}
    (\up{xw}P^{\tau \mu}\ \up{xw}P^{\sigma \nu})
    \Big].
    \nonumber
\end{align}
A similar expression can be derived for the second contraction as 
\hugh{
\begin{align}
\braket{\up{x}\Phi}{\wick{(\c1 a^{\mu})^{\dagger}(\c2 a^{\nu})^{\dagger} \c1 a^{\sigma} \c2 a^{\tau}}&}{\up{w}\Phi}
    =
    \up{xw}\tilde{S} \times 
    \\
    &\Big[
    \hugh{\delta_{0,m}}
    (\up{xw}M^{\sigma \mu}\ \up{xw}M^{\tau \nu})
    + 
    \hugh{\delta_{0,m-1}}
    (\up{xw}M^{\sigma \mu} \ \up{xw}P^{\tau \nu} + \ \up{xw}P^{\sigma \mu} \ \up{xw}M^{\tau \nu} ) 
    + 
    \hugh{\delta_{0,m-2}}
    (\up{xw}P^{\sigma \mu}\ \up{xw}P^{\tau \nu})
    \Big].
    \nonumber
\end{align}
}
%
which, analogously with the orthogonal case, will carry a $-1$ phase factor from the intersection
of the contraction lines.
Combining these two equations, with their associated phase factors, yields the full expression
 for the two-body reduced co-density matrix elements \hugh{with $m$ zero-overlap orbitals} as
\begin{align}
\begin{split}
\braket{^{x}\Phi}{(a^{\mu})^{\dagger} (a^{\nu})^{\dagger} a^{\sigma} a^{\tau}}{^{w}\Phi}
=
\begin{cases}
\up{xw}\tilde{S} \Big(\up{xw}M^{\tau \mu} \ \up{xw}M^{\sigma \nu} -\ \up{xw}M^{\sigma \mu} \ \up{wx}M^{\tau \nu} \Big)
& m=0,
\\
\up{xw}\tilde{S} \Big(\up{xw}P^{\tau \mu}\ \up{xw}M^{\sigma \nu}+\ \up{xw}M^{\tau \mu}\ \up{xw}P^{\sigma \nu} 
 - \up{xw}P^{\sigma \mu}\ \up{xw}M^{\tau \nu} - \up{xw}M^{\sigma \mu}\ \up{xw}P^{\tau \nu}\Big)
& m=1,
\\
\up{xw}\tilde{S} \Big(\up{xw}P^{\tau \mu}\ \up{xw}P^{\sigma \nu} - \up{xw}P^{\sigma \mu}\ \up{xw}P^{\tau \nu}\Big)
& m=2,
\\
0 & m \geq 3.
\end{cases}
\end{split}
\label{eq:twoBodyCoDensity}
\end{align}

\subsection{General Rules for Constructing Matrix Elements}
\label{subsec:GeneralRules}
To simplify the derivation of even longer operator strings, we note that 
the two-body matrix elements in Eq.~\eqref{eq:two-body} can be factorised into the product 
of two fundamental contractions with individual $m_1$ and $m_2$ values under the constraint $m_1 + m_2 = m$, giving
\begin{align}
\begin{split}
    \braket{\up{x}\Phi}{&\wick{(\c2 a^{\mu})^{\dagger}(\c1 a^{\nu})^{\dagger} \c1 a^{\sigma} \c2 a^{\tau}}}{\up{w}\Phi}
    = \up{xw}\tilde{S}
    \sum_{\substack{m_1,m_2 \\ m_1+m_2 = m}}
    \qty( \hugh{\delta_{0,m_1}}\, \up{xw}M^{\tau \mu} + \hugh{\delta_{0,m_1-1}}\, \up{xw}P^{\tau \mu} )
    \qty( \hugh{\delta_{0,m_2}}\, \up{xw}M^{\sigma \nu} + \hugh{\delta_{0,m_2-1}}\, \up{xw}P^{\sigma \nu} ).
\end{split}
\label{eq:factored_contract}
\end{align}
\end{widetext}
\hugh{This factorisation can be extended for a general product of contractions with}
each $m_i$ restricted to the values $0$ or $1$ for the overall product to be non-zero.
We can therefore define an 
intuitive approach for extending Wick's theorem to generalised nonorthogonal matrix elements:
\begin{enumerate}
\item{\label{step:contract}%
    Construct all fully contracted combinations of the operator product with the associated phase factors.}
\item{\label{step:sum_m}%
    For each term, sum every possible way to distribute $m$ zeros among the contractions such that $\sum_i m_i = m$.}
\item{\label{step:factor}%
    For every set of $\{m_i\}$ in each term, construct the relevant contribution 
    as a product of fundamental contractions depending
    on whether each contraction has $m_i = 0$ or $1$.}
\item{\label{step:overlap}
    Multiply the final combined expression with the reduced overlap $\up{xw}\tilde{S}$.}
\end{enumerate}
These rules and contractions therefore allow any matrix element to be evaluated with respect to the asymmetric
Fermi vacuum $\braket{\up{x}\Phi}{\cdots}{\up{w}\Phi}$ for nonorthogonal orbitals,
regardless of whether the many-body  determinants are orthogonal or not.
As a result, our formulation is the most flexible form of Wick's theorem, and reduces
to the previous nonorthogonal\cite{RingBook,Balian1969,Hendekovic1981} or orthogonal\cite{ShavittBook}
variants under suitable restrictions on the MO coefficients.
We refer to this approach as the ``{Extended Nonorthogonal Wick's Theorem}''.
In the following Sections, we will show how these steps can be applied to recover the generalised 
Slater--Condon rules,\cite{MayerBook} and to derive matrix elements between excited configurations with respect to 
the nonorthogonal reference determinants. 

\section{Generalised Slater--Condon Rules}
\label{sec:gsc}

The generalised Slater--Condon rules provide a first-quantised approach for evaluating 
matrix elements between nonorthogonal determinants. 
A detailed description of these rules, their derivation, and their application, can be found in Ref.~\onlinecite{MayerBook}.
However, to the best of our knowledge, the generalised Slater--Condon rules have never previously been derived
though a fully second-quantised framework.
In this Section, we will show how these rules can be recovered using the 
extended nonorthogonal Wick's theorem.

\subsection{One-Body Operators}
\label{subsec:one-body_gsc}

Consider first the one-body operator
\begin{align}
\hat{f} = \sum_{\mu \nu} f_{\mu \nu} (a^{\mu})^{\dagger} a^{\nu},
\end{align}
with the corresponding matrix element
\begin{align}
	\braket{^{x}\Phi}{\hat{f}}{^{w}\Phi} 
	= \sum_{\mu \nu} f_{\mu \nu} \braket{^{x}\Phi}{(a^{\mu})^{\dagger} a^{\nu}}{^{w}\Phi}.
\label{eq:OneBodyExpVal}
\end{align}
Applying the extended nonorthogonal Wick's theorem yields only one non-zero contraction to give
\begin{align}
	\braket{^{x}\Phi}{\hat{f}}{^{w}\Phi} 
	= \sum_{\mu \nu} f_{\mu \nu} \braket{^{x}\Phi}{%
        \wick{(\c1 a^{\mu})^{\dagger} \c1 a^{\nu}}
    }{^{w}\Phi}.
\end{align}
Substituting the fundamental contraction [Eq.~\eqref{eq:fun_contract}] and considering the possible
values of $m$ immediately yields the one-body generalised Slater--Condon rules\cite{MayerBook}
\hugh{in the formed presented in Ref.~\onlinecite{Thom2009} as}
\begin{align}
    \hspace{-5pt}
\braket{^{x}\Phi}{\hat{f}}{^{w}\Phi}  =
\begin{cases}
\up{xw}\tilde{S} \sum_{\mu \nu}  f_{\mu \nu}(\up{xw}W)^{\nu \mu}		& \text{no zeros},
\\[1.5ex]
\up{xw}\tilde{S} \sum_{\mu \nu}  f_{\mu \nu}(\up{xw}P_k)^{\nu \mu}  & \text{one zero ($k$)},
\\[1.5ex]
0								& \text{otherwise.}
\end{cases}
\end{align}
Here, we note that $\up{xw}M^{\nu \mu} = \up{xw}W^{\nu \mu}$
when there are no zero-overlap orbitals, while $\up{xw}P^{\nu \mu}  = \up{xw}P^{\nu \mu}_k$ for
one-zero overlap in orbital $k$.
The matrices $\up{xw}M$ and $\up{xw}P_k$ correspond respectively to the weighted and unweighted
co-density matrices discussed in Ref.~\onlinecite{Thom2009}.

\subsection{Two-Body Operators}
\label{subsec:TwoBodyGenSlatCon}

Next, consider a general two-body operator $\hat{v}$ defined in second-quantisation as 
\begin{align}
\hat{v} = \sum_{\mu \nu \sigma \tau} v_{\mu \nu \tau \sigma} (a^{\mu})^{\dagger} (a^{\nu})^{\dagger} a^{\sigma} a^{\tau}
\end{align}
with the two-electron integrals in the spin-orbital basis defined as
\begin{align}
 v_{\mu \nu \tau \sigma} = \braket{\chi_\mu \chi_\nu}{\hat{v}}{\chi_\tau \chi_\sigma}.
\end{align}
The corresponding matrix element is given by
\begin{align}
\braket{^{x}\Phi}{\hat{v}}{^{w}\Phi} = 
\sum_{\mu \nu \sigma \tau} v_{\mu \nu \tau \sigma}
\braket{^{x}\Phi}{(a^{\mu})^{\dagger} (a^{\nu})^{\dagger} a^{\sigma} a^{\tau}}{^{w}\Phi}.
\end{align}
Recognising the $\braket{^{x}\Phi}{(a^{\mu})^{\dagger} (a^{\nu})^{\dagger} a^{\tau} a^{\sigma}}{^{w}\Phi}$
term as the two-body reduced co-density matrix derived in Eq.~\eqref{eq:twoBodyCoDensity}, 
we immediately recover
\begin{widetext}
\begin{align}
\begin{split}
\braket{^{x}\Phi}{\hat{v}}{^{w}\Phi} = 
\begin{cases}
\up{xw}\tilde{S}\ \sum_{\mu \nu\sigma\tau} v_{\mu \nu \tau \sigma} \Big(\up{xw}M^{\tau \mu} \ \up{xw}M^{\sigma \nu} -\ \up{xw}M^{\sigma \mu} \ \up{xw}M^{\tau \nu} \Big)
& m=0,
\\
\up{xw}\tilde{S}\ \sum_{\mu \nu\sigma\tau} v_{\mu \nu \tau \sigma} \Big(\up{xw}P^{\tau \mu}\ \up{xw}M^{\sigma \nu}+\ \up{xw}M^{\tau \mu}\ \up{xw}P^{\sigma \nu} 
 - \up{xw}P^{\sigma \mu}\ \up{xw}M^{\tau \nu} - \up{xw}M^{\sigma \mu}\ \up{xw}P^{\tau \nu}\Big)
& m=1,
\\
\up{xw}\tilde{S}\ \sum_{\mu \nu\sigma\tau} v_{\mu \nu \tau \sigma} \Big(\up{xw}P^{\sigma \mu}\ \up{xw}P^{\tau \nu} - \up{xw}P^{\tau \mu}\ \up{xw}P^{\sigma \nu}\Big)
& m=2.
\\
0 & m \geq 3,
\end{cases}
\end{split}
\end{align}
Here, note that the $m=1$ terms each contain two terms with the single zero-overlap 
in either the first or second contraction respectively.
Exploiting the symmetry $v_{\mu \nu \tau \sigma} = v_{\nu \mu \sigma \tau}$ and the identity
\begin{align}
    \up{xw}P^{\sigma \mu}\ ^{xw}P^{\tau \nu}  - \up{xw}P^{\tau \mu}\ ^{xw}P^{\sigma \nu}
    =
    \sum_{k_1 \neq k_2}
    \qty(
    \up{xw}P_{k_1}^{\sigma \mu}\ ^{xw}P_{k_2}^{\tau \nu} - \up{xw}P_{k_1}^{\tau \mu}\ ^{xw}P_{k_2}^{\sigma \nu}
    )
\end{align}
then allows the two-body generalised Slater--Condon rules\cite{MayerBook} to be recovered as
\begin{align}
\begin{split}
\braket{^{x}\Phi}{\hat{v}}{^{w}\Phi}  =
\begin{cases}
\up{xw}\tilde{S}\ \sum_{\mu \nu\sigma\tau} v_{\mu \nu \tau \sigma} \Big(\up{xw}W^{\tau \mu} \ \up{xw}W^{\sigma \nu} - \up{xw}W^{\sigma \mu} \ \up{xw}W^{\tau \nu} \Big)
& \text{no zeros},
\\[1.5ex]
2\ \up{xw}\tilde{S}\ \sum_{\mu \nu\sigma\tau} v_{\mu \nu \tau \sigma} \Big(\up{xw}P_k^{\tau \mu}\ \up{xw}W^{\sigma \nu} - \up{xw}P_k^{\sigma \mu}\ \up{xw}W^{\tau \nu}\Big)
& \text{one zero ($k$)},
\\[1.5ex]
2\ \up{xw}\tilde{S}\  \sum_{\mu \nu\sigma\tau} v_{\mu \nu \tau \sigma}  \Big(\up{xw}P_{k_1}^{\tau \mu}\ \up{xw}P_{k_2}^{\sigma \nu} - \up{xw}P_{k_1}^{\sigma \mu}\ \up{xw}P_{k_2}^{\tau \nu}\Big)
& \text{two zeros ($k_1,k_2$)},
\\[1.5ex]
0 & \text{otherwise.}
\end{cases}
\end{split}
\end{align}
Note that, for the $m=1$ case with one zero-overlap in orbital $k$, we have exploited the identities 
$\up{xw}P_k^{\tau \mu}\ \up{xx}P_k^{\sigma \nu}  = \up{xw}P_k^{\sigma \mu}\ \up{xx}P_k^{\mu \nu} $
and
$\up{xw}P_k^{\tau \mu}\ \up{xw}P_k^{\sigma \nu} =\up{xw}P_k^{\sigma \mu}\ \up{xw}P_k^{\mu \nu} $
to introduce the simplification $\up{xw}M \rightarrow  \up{xw}W$.
\end{widetext}

\section{Matrix Elements for Excited Configurations}
\label{sec:ExcitedConfig}

While re-deriving the generalised Slater--Condon rules provides an important verification of the extended nonorthogonal 
Wick's theorem, it does not provide any computational advantage over the original framework.
On the contrary, the primary focus of our new framework involves 
deriving matrix elements between excited configurations from a pair of nonorthogonal determinants
\eg,  $\braket{\up{x}\Phi_{ij...}^{ab...}}{\cdots}{\up{w}\Phi_{kl...}^{cd...}}$.
Terms of this form arise in perturbative corrections to NOCI,\cite{Yost2013,Yost2016,Yost2019,Burton2020}
the NOCI-CIS approach for core excitations,\cite{Oosterbaan2018,Oosterbaan2019}
and the evaluation of $\langle S^2 \rangle$ coupling terms in NOCI expansions.\cite{Sundstrom2014}
Furthermore, these nonorthogonal matrix elements will be required to
evaluate inter-state coupling elements between orbital-optimised excited-state wave functions identified
using excited-state mean-field theory\cite{Shea2018} or state-specific complete active space SCF.\cite{Tran2019}

Until now, evaluating these matrix elements has required the
direct application of the generalised Slater--Condon rules to each pair of excitations.
This approach leads to significant computational costs associated with the biorthogonalisation of 
the excited determinants, \hugh{which scales as $\Or{\Ne^3}$ each time.}
In this Section, we show how the extended nonorthogonal Wick's theorem allows these matrix
elements to be evaluated using only biorthogonalised reference determinants. 
\hugh{When large numbers of coupling elements are required, avoiding the 
biorthogonalisation of these excited configurations significantly reduces the overall computational cost.}
\hugh{Furthermore, the cost of certain nonorthogonal matrix elements (including all one-body operators)
becomes independent of the number of electrons or basis functions if additional intermediate matrix elements
are computed for each pair of reference determinants.}

\hugh{In this Section, we describe how the extended nonorthogonal Wick's theorem can be applied to 
matrix elements between excited configurations.
First, we discuss how the fundamental contractions described in Section~\ref{subsec:FundamentalContraction} 
can be modified to an MO-based form that makes excited configurations easier to handle.
We then illustrate the derivation of certain overlap, one-body, and two-body matrix elements between nonorthogonal excited 
configurations.
The resulting expressions are entirely generalised for any pair of nonorthogonal reference determinants and can
significantly reduce the computational scaling compared to a na\"ive application of the generalised Slater--Condon rules.}

\subsection{Asymmetric Representation}
\label{subsec:AsymmRepresentation}
Evaluating matrix elements between two excited determinants will require the evaluation of the asymmetric
contractions $\wick{\up{x}\c1 b_{p}^{\dagger} \up{w} \c1 b_{q}}$ with respect
to the asymmetric Fermi vacuum $\braket{\up{x}\Phi}{\cdots}{\up{w}\Phi}$.
Expanding the molecular orbital creation and annihilation operators using Eq.~\eqref{eq:MObOperators} yields 
\begin{subequations}
\begin{align}
\wick{\up{x}\c1 b_{p}^{\dagger} \up{w} \c1 b_{q}} 
= \sum_{\mu \nu \sigma \tau} (\up{w}C^{*})^{\cdot \mu}_{q \cdot}\, g_{\mu \sigma}\,\wick{(\c1 a^{\tau})^{\dagger} (\c1 a^{\sigma})}\, g_{\tau \nu}\, (\up{x}C)^{\nu \cdot}_{\cdot p},
\\
\wick{\up{x}\c1 b_{p} \up{w} \c1 b_{q}^{\dagger}} 
= \sum_{\mu \nu \sigma \tau} (\up{x}C^{*})^{\cdot \mu}_{p \cdot}\, g_{\mu \sigma}\,\wick{(\c1 a^{\sigma}) (\c1 a^{\tau})^{\dagger} }\, g_{\tau \nu}\, (\up{w}C)^{\nu \cdot}_{\cdot q}.
\end{align}
\label{eq:barePair}
\end{subequations}
Introducing the fundamental contractions \hugh{for $\wick{(\c1a^{\mu})^{\dagger} \c1a^{\nu}}$ and  
$\wick{(\c1a^{\mu})^{\dagger} \c1a^{\nu}}$ defined in} 
Eqs.~\eqref{eq:fun_contract} and \eqref{eq:fun_contract2} respectively then reduces these expressions to different
forms depending on the number of zero-overlaps $m_k$ associated with the contraction, 
\begin{subequations}
\begin{align}
\wick{\up{x}\c1 b_{p}^{\dagger} \up{w} \c1 b_{q}} 
&=
\begin{cases}
\up{wx}X_{qp} 	  & m_k=0,
\\
\up{wx}\bX_{qp}   & m_k=1,
\\
0					  & m_k >1,
\end{cases}
\label{eq:funPairContract}
\\
\wick{\up{x}\c1 b_{p} \up{w} \c1 b_{q}^{\dagger}} 
&=
\begin{cases}
\up{xw}Y_{pq} 	  & m_k=0,
\\
\up{xw}\bY_{pq}   & m_k=1,
\\
0					  & m_k >1.
\end{cases}
\label{eq:funPairContract2}
\end{align}
\end{subequations}
Here, we have defined the ``screened'' overlap terms
\begin{subequations}
\begin{align}
\up{wx}X_{qp} &= 
\sum_{\mu \nu \sigma \tau} (\up{w}C^{*})^{\cdot \mu}_{q \cdot}\, g_{\mu \sigma}(\up{xw}M)^{\sigma \tau}\, g_{\tau \nu}\, (\up{x}C)^{\nu \cdot}_{\cdot p},
\label{eq:funContractLeft}
\\
\up{wx}\bX_{qp} &= 
\sum_{\mu \nu \sigma \tau} (\up{w}C^{*})^{\cdot \mu}_{q \cdot}\, g_{\mu \sigma}(\up{xw}P)^{\sigma \tau}\, g_{\tau \nu}\, (\up{x}C)^{\nu \cdot}_{\cdot p},
\label{eq:funContractRight}
\end{align}
\end{subequations}
and 
\begin{subequations}
\begin{align}
\up{xw}Y_{pq} &=  - \qty(\up{xw}X_{pq} - \up{xw}S_{pq})
\\
\up{xw}\bY_{pq} &= -\up{xw} \bX_{pq},
\end{align}
\end{subequations}
with $\up{xw}M$ and $\up{xw}P$ defined in Eqs.~\eqref{eq:CoDensitySum} and \eqref{eq:Msum} respectively, 
and the overlap element
\begin{align}
 \up{wx}S_{pq} = \sum_{\mu \nu} (\up{w}C^{*})^{\cdot \mu}_{p \cdot}\,g_{\mu \nu}\, (\up{x}C)^{\nu \cdot}_{\cdot q}. 
\end{align}
\hugh{Although the indexing notation used in Eq.~\eqref{eq:funContractLeft} may seem counterintuitive, we find that it helps
to keep track of the bra and ket orbital coefficients in the screened overlap terms.}

Crucially, the orbital coefficients used to evaluate these contractions do not need to 
be the same as those used to evaluate the $\up{xw}M$ and $\up{xw}P$ matrices.
This feature is particularly advantageous as the excited configurations can be 
represented in terms of the original orbital basis while the
$\up{xw}M$ and $\up{xw}P$ matrices are evaluated in the biorthogonal basis.
As a result, only the reference determinants need to be biorthogonalised, and the remaining matrix elements 
are evaluated in terms of these screened overlap terms.
Furthermore, the screened overlap elements are themselves one-body matrix elements that can be 
computed once for a given pair of determinants
and stored, before being combined to evaluate more complicated matrix elements.
\hugh{With $\Nref$ determinants, the total cost of computing these intermediates therefore scales
as $\Or{\Nref^2\, \Nbas^3}$.}

To take full advantage of these asymmetric contractions, the one- and two-body
operators can also be represented in terms of one set of molecular orbitals as
\begin{subequations}
\begin{align}
\hat{f} &= \sum_{pq} \up{x}f_{pq} \up{x}b_{p}^{\dagger} \up{x}b_{q}
\label{eq:one_body}
\\
\hat{v} &= \sum_{pqrs} \up{x}v_{pqrs} \up{x}b_{p}^{\dagger} \up{x}b_{q}^{\dagger} \up{x}b_{s} \up{x}b_{r},
\label{eq:x_two_body}
\end{align}
\end{subequations}
where we have defined the transformed matrix elements
\begin{align}
\up{x}f_{pq} = \sum_{\mu \nu} (\up{x}C^{*})^{\cdot \mu}_{p \cdot}\,f_{\mu \nu}\,(\up{x}C)^{\nu \cdot}_{\cdot q}
\end{align}
and
\begin{align}
\up{x}v_{pqrs} = \sum_{\mu \nu \sigma \tau} (\up{x}C^{*})^{\cdot \mu}_{p \cdot} (\up{x}C^{*})^{\cdot \nu}_{q \cdot}\,
v_{\mu \nu \sigma \tau}\,
(\up{x}C)^{\sigma \cdot}_{\cdot r} (\up{x}C)^{\tau \cdot}_{\cdot s}.
\end{align}
Evaluating nonorthogonal matrix elements through the extended
nonorthogonal Wick's theorem then proceeds using similar steps outlined in Section~\ref{subsec:GeneralRules}:
\hugh{
\begin{enumerate}
\item{\label{step:contract2}%
    Assemble all fully contracted combinations of the asymmetric operator and excitation operator product
    and compute the corresponding phase factors.}
\item{\label{step:sum_m2}%
    For each term, sum every possible way to distribute $m$ zeros among the contractions such that $\sum_i m_i = m$.}
\item{\label{step:factor2}%
    For every set of $\{m_i\}$ in each term, construct the relevant contribution 
    as a product of fundamental contractions defined in Eqs.~\eqref{eq:funPairContract} and \eqref{eq:funPairContract2}
    depending on whether each contraction has $m_i = 0$ or $1$.}
\item{\label{step:overla2}
    Multiply the combined expression with the reduced overlap $\up{xw}\tilde{S}$ of the reference determinants.}
\end{enumerate}}
Note that the number of zero-overlap orbitals in these expressions corresponds to the 
biorthogonalised reference determinants, not the excited configurations.

In the following Sections, we will illustrate this process through a series of typical nonorthogonal matrix elements.
As we shall see below, evaluating the sum of every combination of zero-overlap indices assigned to each
contraction quickly leads to complicated equations. 
However, we can derive the general structure of a matrix element for the case with no zero-overlap orbitals $m=0$, 
and then recover the forms for different values of $m$ by distributing the zero-overlap indices over each 
fundamental contraction. 
We will therefore focus on the parent equation with $m=0$, which we refer to as the ``canonical form'' 
\hugh{and denote using the notation $\expval*{\cdots}_0$}.
Furthermore, any matrix element with more zero-overlap orbitals than the total number of contractions must 
be strictly zero.

\subsection{Overlap Terms}
\label{subsec:OverlapTerms}

First, we consider the overlap element between excited determinants.
The simplest overlap matrix element involves only a single excitation $\brkt{^{x}\Phi}{^{w}\Phi_i^a}$.
For $m$ zero-overlap orbitals in the reference determinants, this single excitation matrix element can be 
identified using one contraction as
\begin{align}
\begin{split}
\brkt{^{x}\Phi}{^{w}\Phi_i^a} 
&= 
\braket{^{x}\Phi}{\wick{\up{w} \c1 b_a^{\dagger} \up{w} \c1 b_i^{\vphantom{\dagger}}}}{^{w}\Phi}
=
\begin{cases}
\up{xw}\tilde{S}\ \up{ww}X_{ia} 	  & m=0,
\\
\up{xw}\tilde{S}\ \up{ww}\bar{X}_{ia}   & m=1,
\\
0					  & m >1.
\end{cases}
\end{split}
\end{align}
Here, the canonical form is simply
\begin{align}
\brkt{^{x}\Phi}{^{w}\Phi_i^a}_0 = \up{xw}\tilde{S}\ \up{ww}X_{ia} ,
\end{align}
and there is only one way to assign one zero-overlap orbital to the single contraction.
Note that the reduced overlap between the \emph{reference} determinants remains a prefactor
for the overall matrix elements.

Next, the overlap of two single excitations $\brkt{^{x}\Phi_i^a}{^{w}\Phi_j^b}$ can be evaluated as
\begin{widetext}
\begin{align}
\begin{split}
\brkt{^{x}\Phi_i^a}{^{w}\Phi_j^b} 
&= 
\braket{^{x}\Phi}{\wick{\up{x} \c1 b_i^{\dagger} \up{x} \c1 b_a^{\vphantom{\dagger}} \up{w} \c1 b_b^{\dagger} \up{w} \c1 b_j^{\vphantom{\dagger}}}}{^{w}\Phi}
+
\braket{^{x}\Phi}{\wick{\up{x} \c2 b_i^{\dagger} \up{x} \c1 b_a^{\vphantom{\dagger}} \up{w} \c1 b_b^{\dagger} \up{w} \c2 b_j^{\vphantom{\dagger}}}}{^{w}\Phi}
\\
&=
\begin{cases}
\up{xw}\tilde{S}\qty( \up{ww}X_{jb}  \up{xx}X_{ai} + \up{wx}X_{ji} \up{xw}Y_{ab} 	)  
& m=0,
\\
\up{xw}\tilde{S}\qty( \up{ww}X_{jb}  \up{xx}\bX_{ai} + \up{wx}X_{ji} \up{xw}\bY_{ab} 	  + \up{ww}\bX_{jb}  \up{xx}X_{ai} + \up{wx}\bX_{ji} \up{xw}Y_{ab} 	)  
& m=1,
\\
\up{xw}\tilde{S}\qty( 
\up{ww}\bX_{jb}  \up{xx}\bX_{ai} +  \up{wx}\bX_{ji} \up{xw}\bY_{ab})   
& m=2,
\\
0					  
& m>2.
\end{cases}
\end{split}
\label{eq:sing_sing_ov}
\end{align}
The canonical form for this element is the $m=0$ term 
\begin{align}
\brkt{^{x}\Phi_i^a}{^{w}\Phi_j^b}_0 = \up{xw}\tilde{S}\qty( \up{ww}X_{jb}  \up{xx}X_{ai} + \up{wx}X_{ji} \up{xw}Y_{ab} 	) .
\end{align}
The $m=1$ term is recovered by taking the sum of the two different ways to distribute 
one zero-overlap orbital to the two contractions, while there is only one way to distribute two zero-overlap
orbitals for the $m=2$ term.

As a third example, consider the double excitation overlap $\brkt{^{x}\Phi}{^{w}\Phi_{ij}^{ab}}$, which 
can be evaluated as
\begin{align}
\begin{split}
\brkt{^{x}\Phi}{^{w}\Phi_{ij}^{ab}}
&= 
\braket{^{x}\Phi}{\wick{\up{w} \c2 b_a^{\dagger} \up{w} \c1 b_b^{\dagger} \up{w} \c1 b_j \up{w} \c2 b_i^{\vphantom{\dagger}}}}{^{w}\Phi}
+ 
\braket{^{x}\Phi}{\wick{\up{w} \c1 b_a^{\dagger} \up{w} \c2 b_b^{\dagger} \up{w} \c1 b_j \up{w} \c2 b_i^{\vphantom{\dagger}}}}{^{w}\Phi}
\\
&=
\begin{cases}
\up{xw}\tilde{S}\qty(\up{ww}X_{ia} \up{ww}X_{jb} - \up{ww}X_{ib} \up{ww}X_{ja})	 
& m=0,
\\
\up{xw}\tilde{S}\qty(\up{ww}X_{ia} \up{ww}\bar{X}_{jb} - \up{ww}X_{ib} \up{ww}\bar{X}_{ja} + \up{ww}\bar{X}_{ia} \up{ww}X_{jb} - \up{ww}\bar{X}_{ib} \up{ww}X_{ja})  
& m=1,
\\
\up{xw}\tilde{S}\qty(\up{ww}\bar{X}_{ia} \up{ww}\bar{X}_{jb} - \up{ww}\bar{X}_{ib} \up{ww}\bar{X}_{ja})  
& m=2,
\\
0					  
& m >2.
\end{cases}
\end{split}
\end{align}
\end{widetext}
In this case, the minus sign arises from the intrinsic $-1$ phase that results from the intersection of the contraction lines in
$\braket{^{x}\Phi}{\wick{\up{w} \c1 b_a^{\dagger} \up{w} \c2 b_b^{\dagger} \up{w} \c1 b_j \up{w} \c2 b_i^{\vphantom{\dagger}}}}{^{w}\Phi}$.
The relevant canonical form of this matrix element is 
\begin{align}
\brkt{^{x}\Phi}{^{w}\Phi_{ij}^{ab}}_0 = \up{xw}\tilde{S}\qty(\up{ww}X_{ia} \up{ww}X_{jb} - \up{ww}X_{ib} \up{ww}X_{ja}),	 
\end{align}
from which the terms for $m=1$ and $m=2$ can be derived.
Notably, this form of $\brkt{^{x}\Phi}{^{w}\Phi_{ij}^{ab}}$ has previously be derived for $m=0$ in Refs.~\onlinecite{Yost2013} and \onlinecite{Nite2019b}, but 
our derivation generalises for any number of zero-overlap orbitals.

\subsection{One-Body Operators}
\label{subsec:OneBody}
We now consider matrix elements for one-body operators of the form given in Eq.~\eqref{eq:one_body}.
To further simplify the subsequent expressions, we can introduce intermediate matrices that account for 
partial contraction with the one-body operator.
In particular, we introduce the partially contracted intermediate terms
\begin{subequations}
\begin{align}
F_0
&
= \sum_{pq} \up{x}f_{pq} \up{xx}X_{qp}
\\
\up{yw}[XFX]_{rs}
&
= \sum_{pq} \up{yx}X_{rp} \up{x}f_{pq} \up{xw}X_{qs}
\\
\up{yw}[XFY]_{rs}
&
= \sum_{pq} \up{yx}X_{rp} \up{x}f_{pq} \up{xw}Y_{qs}
\\
\up{yw}[YFX]_{rs}
&
= \sum_{pq} \up{yx}Y_{rp} \up{x}f_{pq} \up{xw}X_{qs}
\\
\up{yw}[YFY]_{rs}
&
= \sum_{pq} \up{yx}Y_{rp} \up{x}f_{pq} \up{xw}Y_{qs}
\end{align}
\label{eq:oneBodyInter}
\end{subequations}
Note that the individual contraction in the $F_0$ term, indicated by the $X$ matrix, may
correspond to a zero-overlap orbital pair, as indicated by the notation $\bar{F}_0$.
The $\up{yw}[XFX]_{rs}$ (and similar) terms correspond to two contractions, and can thus be assigned two 
zero overlap orbitals. 
The different possibilities of assigning these zero-overlap contractions is denoted as 
$\up{yw}[\bX FX]_{rs}$ or $\up{yw}[X F \bX]_{rs}$, and similarly for terms involving the $Y$ contraction.
Crucially, these intermediate values also correspond to orbital pairs, and can be precomputed
once for each pair of reference determinants, \hugh{leading to a one-off computational cost that scales
as $\Or{\Nref^2\,n^3}$}.
As a result, the summation over the $p,q$ indices is avoided for the subsequent evaluation of matrix
elements between excited determinants, \hugh{and the computational scaling of these one-body terms is the 
same as the overlap matrix elements}.

To illustrate the application of this approach, we 
 take the simplest one-body matrix element $\braket{\up{x}\Phi}{\hat{f}}{\up{w}\Phi_i^a}$.
For $m$ zero-overlap orbitals in the reference determinants, this matrix element can be expanded as
\begin{align}
\begin{split}
\braket{\up{x}\Phi}{&\hat{f}}{\up{w}\Phi_i^a}
= 
\sum_{pq} \up{x}f_{pq} \times 
\\
&\qty(
\braket{\up{x}\Phi}{ \wick{\up{x} \c1 b_p^{\dagger} \up{x} \c1 b_q \up{w} \c1 b_a^{\dagger} \up{w} \c1 b_i^{\vphantom{\dagger}} } }{\up{w}\Phi}
+
\braket{\up{x}\Phi}{ \wick{\up{x} \c2 b_p^{\dagger} \up{x} \c1 b_q \up{w} \c1 b_a^{\dagger} \up{w} \c2 b_i^{\vphantom{\dagger}} } }{\up{w}\Phi}
).
\end{split}
\end{align}
Using the contractions defined in Eqs.~\eqref{eq:funPairContract} and \eqref{eq:funPairContract2}, and the intermediate 
terms defined in Eq.~\eqref{eq:oneBodyInter}, the canonical form for $m=0$ is given as
\begin{align}
\canon{\up{x}\Phi}{\hat{f}}{\up{w}\Phi_i^a} 
=
\up{xw}\tilde{S} \qty(F_0 \up{ww}X_{ia} + \up{ww}[XFY]_{ia}).
\end{align}
If the $F_0$, $\up{ww}X_{ia} $, and $\up{ww}[XFY]_{ia}$ are precomputed and stored once for 
the reference determinants, the subsequent cost of evaluating $\braket{\up{x}\Phi}{\hat{f}}{\up{w}\Phi_i^a}$ is
\emph{independent} of the number of electrons or basis functions as scales as $\Or{1}$.
From this canonical form, expressions for $m=1$ and $2$ (or higher) can be identified as 
\begin{widetext}
\begin{align}
\begin{split}
\braket{\up{x}\Phi}{\hat{f}}{\up{w}\Phi_i^a}
&= 
\begin{cases}
\up{xw}\tilde{S} \qty(F_0 \up{ww}X_{ia} + \up{ww}[XFY]_{ia})
& m=0,
\\
\up{xw}\tilde{S} \qty(\bar{F}_0 \up{ww}X_{ia} + \up{ww}[\bX FY]_{ia} + F_0 \up{ww}\bX_{ia} + \up{ww}[X F \bY]_{ia} )
& m=1,
\\
\up{xw}\tilde{S} \qty(\bar{F}_0 \up{ww}\bX_{ia} + \up{ww}[\bX F\bY]_{ia} )
& m=2,
\\
0					  & m > 2.
\end{cases}
\end{split}
\end{align}

Next, consider the coupling of two single excitations $\braket{\up{x}\Phi_i^a}{\hat{f}}{\up{w}\Phi_j^b}$, corresponding to the fully contracted terms
\begin{align}
\begin{split}
\braket{\up{x}\Phi_i^a}{&\hat{f}}{\up{w}\Phi_j^b}
= 
\sum_{pq} \up{x}f_{pq} \times
\\
\Big(
&
\braket{\up{x}\Phi}{ \wick{\up{x} \c1 b_i^{\dagger} \up{x} \c1 b_a \up{x} \c1 b_p^{\dagger} \up{x} \c1b_q \up{w} \c1 b_b^{\dagger} \up{w} \c1 b_j } }{\up{w}\Phi}
+
\hugh{\braket{\up{x}\Phi}{ \wick{\up{x} \c3 b_i^{\dagger} \up{x} \c2 b_a \up{x} \c1 b_p^{\dagger} \up{x} \c1 b_q \up{w} \c2 b_b^{\dagger} \up{w} \c3 b_j } }{\up{w}\Phi}}
+
\braket{\up{x}\Phi}{ \wick{\up{x} \c1 b_i^{\dagger} \up{x} \c1 b_a \up{x} \c2 b_p^{\dagger} \up{x} \c1 b_q \up{w} \c1 b_b^{\dagger} \up{w} \c2 b_j } }{\up{w}\Phi}
+
\\
&+
\braket{\up{x}\Phi}{ \wick{\up{x} \c2 b_i^{\dagger} \up{x} \c1 b_a \up{x} \c1 b_p^{\dagger} \up{x} \c2 b_q \up{w} \c1 b_b^{\dagger} \up{w} \c1 b_j } }{\up{w}\Phi}
+
\braket{\up{x}\Phi}{ \wick{\up{x} \c2 b_i^{\dagger} \up{x} \c1 b_a \up{x} \c1 b_p^{\dagger} \up{x} \c1 b_q \up{w} \c1 b_b^{\dagger} \up{w} \c2 b_j } }{\up{w}\Phi}
+
\braket{\up{x}\Phi}{ \wick{\up{x} \c1 b_i^{\dagger} \up{x} \c2 b_a \up{x} \c3 b_p^{\dagger} \up{x} \c1 b_q \up{w} \c2b_b^{\dagger} \up{w} \c3 b_j } }{\up{w}\Phi}
\Big).
\end{split}
\label{eq:fockSingSing}
\end{align}
In this case, the canonical form is given by
\begin{align}
\begin{split}
\canon{\up{x}\Phi_i^a}{\hat{f}}{\up{w}\Phi_j^b}
=
\up{xw}\tilde{S} \Big( & F_0 \up{xx}X_{ai} \up{ww}X_{jb} + F_0  \up{wx}X_{ji} \up{xw}Y_{ab}
\\&+ \up{ww}[XFY]_{jb} \up{xx}X_{ai} + \up{xx}[YFX]_{ai} \up{ww}X_{jb} 
+ \up{xw}[YFY]_{ab} \up{wx}X_{ji} 
- \up{wx}[XFX]_{ji} \up{xw}Y_{ab} \Big).
\end{split}
\end{align}
Again, the form for $m$ zero-overlap orbitals can be recovered as the sum of every way to
 distribute the zero-overlap contractions over the $X$, $Y$, or $F_0$ terms in each product.
We omit the explicit form of these $m>0$ expressions to maintain brevity.

Finally, consider the one-body coupling of a reference determinant and a double excitation
\begin{align}
\begin{split}
\braket{\up{x}\Phi}{&\hat{f}}{\up{w}\Phi_{ij}^{ab}}
= 
\sum_{pq} \up{x}f_{pq} \times
\\
\Big(
&
\braket{\up{x}\Phi}{ \wick{\up{x} \c1 b_p^{\dagger} \up{x} \c1 b_q \up{w} \c2 b_a^{\dagger} \up{w} \c1b_b^{\dagger}  \up{w} \c1 b_j\up{w} \c2 b_i } }{\up{w}\Phi}
+
\braket{\up{x}\Phi}{ \wick{\up{x} \c1 b_p^{\dagger} \up{x} \c1 b_q \up{w} \c1 b_a^{\dagger} \up{w} \c2b_b^{\dagger}  \up{w} \c1 b_j\up{w} \c2 b_i } }{\up{w}\Phi}
+
\braket{\up{x}\Phi}{ \wick{\up{x} \c2 b_p^{\dagger} \up{x} \c1 b_q \up{w} \c1 b_a^{\dagger} \up{w} \c1b_b^{\dagger}  \up{w} \c2 b_j\up{w} \c1 b_i } }{\up{w}\Phi}
\\
&
+
\braket{\up{x}\Phi}{ \wick{\up{x} \c2 b_p^{\dagger} \up{x} \c1 b_q \up{w} \c3 b_a^{\dagger} \up{w} \c1b_b^{\dagger}  \up{w} \c2 b_j\up{w} \c3 b_i } }{\up{w}\Phi}
+
\braket{\up{x}\Phi}{ \wick{\up{x} \c2 b_p^{\dagger} \up{x} \c1 b_q \up{w} \c1 b_a^{\dagger} \up{w} \c1b_b^{\dagger}  \up{w} \c1 b_j\up{w} \c2 b_i } }{\up{w}\Phi}
+
\braket{\up{x}\Phi}{ \wick{\up{x} \c3 b_p^{\dagger} \up{x} \c1 b_q \up{w} \c2 b_a^{\dagger} \up{w} \c1b_b^{\dagger}  \up{w} \c2 b_j\up{w} \c3 b_i } }{\up{w}\Phi}
),
\end{split}
\label{eq:fockRefDoub}
\end{align}
The corresponding canonical form for $m=0$ is 
\begin{align}
\begin{split}
\canon{\up{x}\Phi}{\hat{f}}{\up{w}\hugh{\Phi_{ij}^{ab}}}
= 
\up{xw}\tilde{S} \Big( 
&F_0 \up{ww}X_{ia} \up{ww}X_{jb} - F_0 \up{ww}X_{ib} \up{ww}X_{ja}
\\
&+ \up{ww}[XFY]_{ia} \up{ww}X_{jb} + \up{ww}[XFY]_{jb} \up{ww}X_{ia} - \up{ww}[XFY]_{ja} \up{ww}X_{ib} - \up{ww}[XFY]_{ib} \up{ww}X_{ja}
\Big),
\end{split}
\end{align}
from which expressions for $m>0$ can be obtained as described above.
\end{widetext}

\subsection{Two-body Operators}

Finally, we consider matrix elements for two-body operators, with the general form given 
in Eq.~\eqref{eq:x_two_body}. 
Again, we can define intermediate matrices in the orbital basis that can be 
pre-computed for a given pair of nonorthogonal determinants.
First, we introduce analogues of the Coulomb and exchange matrices, respectively defined as
\begin{align}
\begin{split}
\up{x}J_{pr} &= \sum_{qs} \up{x}v_{pqrs} \up{xx}X_{sq} 
\\
\up{x}K_{ps} &= \sum_{qr} \up{x}v_{pqrs} \up{xx}X_{rq}. 
\end{split}
\label{eq:JKmats}
\end{align}
Partially contracted intermediate matrices can then be defined as \eg,
\begin{subequations}
\begin{align}
\up{yw}[X(J-K)X]_{rs}
&
= \sum_{pq} \up{yx}X_{rp} (\up{x}J- \up{x}K)_{pq} \up{xw}X_{qs}
\\
\up{yw}[Y(J-K)X]_{rs}
&
= \sum_{pq} \up{yx}Y_{rp} (\up{x}J- \up{x}K)_{pq} \up{xw}X_{qs}
\\
\up{yw}[X(J-K)Y]_{rs}
&
= \sum_{pq} \up{yx}X_{rp} (\up{x}J- \up{x}K)_{pq} \up{xw}Y_{qs}
\\
\up{yw}[Y(J-K)Y]_{rs}
&
= \sum_{pq} \up{yx}Y_{rp} (\up{x}J- \up{x}K)_{pq} \up{xw}Y_{qs},
\end{align}
\label{eq:twoBodyInter}
\end{subequations}
with the constant value
\begin{align}
V_0 = \sum_{pqrs} \up{x}v_{pqrs} \qty(\up{xx}X_{sq} \up{xx}X_{rp} - \up{xx}X_{rq} \up{xx}X_{sp}).
\label{eq:twoBodyConst}
\end{align}
\hugh{The one-off computational cost of evaluating each of these intermediate matrices is dominated by 
the cost of evaluating the Eqs.~\eqref{eq:JKmats} and \eqref{eq:twoBodyConst} for each reference determinant, 
and so the overall scaling is $\Or{\Nref\, \Nbas^4}$.}

Like the one-body intermediate matrices, when the zero-overlap orbitals are 
distributed among the contractions, these zeros may be independently assigned to any of the $X$, $Y$, or 
$(J-K)$ terms, as each of these contain one contraction.
For example, with $m=1$, one must consider the sum of three terms given as \eg,
\begin{align}
\begin{split}
\up{yw}[\bX(J-K)X]_{rs} + \up{yw}[X&(\bJ-\bK)X]_{rs} 
\\
&+ \up{yw}[X(J-K)\bX]_{rs},
\end{split}
\end{align}
where 
\begin{subequations}
\begin{align}
\up{x}\bJ_{pr} &= \sum_{qs} \up{x}v_{pqrs} \up{xx}\bX_{sq} 
\\
\up{x}\bK_{ps} &= \sum_{qr} \up{x}v_{pqrs} \up{xx}\bX_{rq}. 
\end{align}
\end{subequations}
On the contrary, the $V_0$ constant term corresponds to two contractions and can be assigned two
zero-overlap orbitals. 
Noting the symmetry $\up{x}v_{pqrs} = \up{x}v_{qpsr}$, we denote these possibilities as 
\begin{subequations}
\begin{align}
\bar{V}_0 &= 2 \sum_{pqrs} \up{x}v_{pqrs} \qty(\up{xx}\bX_{sq} \up{xx}X_{rp} - \up{xx}\bX_{rq} \up{xx}X_{sp})
\label{eq:V0bar}
\\
\bar{\bar{V}}_0 &=  \sum_{pqrs} \up{x}v_{pqrs} \qty(\up{xx}\bX_{sq} \up{xx}\bX_{rp} - \up{xx}\bX_{rq} \up{xx}\bX_{sp}).
\end{align}
\end{subequations}
\hugh{Here, the factor of two in Eq.~\eqref{eq:V0bar} arises because the zero-overlap orbital can be assigned to either
the first or second contraction to give the same result.}

To illustrate the application of this approach for excited configurations,
we first consider the two-body matrix element 
\begin{widetext}
\begin{align}
\begin{split}
\braket{\up{x}\Phi}{\hat{v}}{\up{w}\Phi_i^a}
= 
&\sum_{pqrs} \up{x}v_{pqrs} \times
\\ \Big(
&\braket{\up{x}\Phi}{ \wick{\up{x} \c2 b_p^{\dagger} \up{x} \c1 b^{\dagger}_q \up{x} \c1 b_s \up{x} \c2 b_r  \up{w} \c1 b_a^{\dagger} \up{w} \c1 b_i^{\vphantom{\dagger}} } }{\up{w}\Phi}
+
\braket{\up{x}\Phi}{ \wick{\up{x} \c1 b_p^{\dagger} \up{x} \c2 b^{\dagger}_q \up{x} \c1 b_s \up{x} \c2 b_r  \up{w} \c1 b_a^{\dagger} \up{w} \c1 b_i^{\vphantom{\dagger}} } }{\up{w}\Phi}
+
\braket{\up{x}\Phi}{ \wick{\up{x} \c2 b_p^{\dagger} \up{x} \c1 b^{\dagger}_q \up{x} \c1 b_s \up{x} \c1 b_r  \up{w} \c1 b_a^{\dagger} \up{w} \c2 b_i^{\vphantom{\dagger}} } }{\up{w}\Phi}
\\
&+
\braket{\up{x}\Phi}{ \wick{\up{x} \c3 b_p^{\dagger} \up{x} \c1 b^{\dagger}_q \up{x} \c2 b_s \up{x} \c1 b_r  \up{w} \c2 b_a^{\dagger} \up{w} \c3 b_i^{\vphantom{\dagger}} } }{\up{w}\Phi}
+
\braket{\up{x}\Phi}{ \wick{\up{x} \c1 b_p^{\dagger} \up{x} \c2 b^{\dagger}_q \up{x} \c1 b_s \up{x} \c1 b_r  \up{w} \c1 b_a^{\dagger} \up{w} \c2 b_i^{\vphantom{\dagger}} } }{\up{w}\Phi}
+
\braket{\up{x}\Phi}{ \wick{\up{x} \c1 b_p^{\dagger} \up{x} \c3 b^{\dagger}_q \up{x} \c2 b_s \up{x} \c1 b_r  \up{w} \c2 b_a^{\dagger} \up{w} \c3 b_i^{\vphantom{\dagger}} } }{\up{w}\Phi}
\Big).
\end{split}
\end{align}
Combining each contraction, and exploiting the intermediates in Eqs.~\eqref{eq:twoBodyInter} and \eqref{eq:twoBodyConst}
yields the canonical form ($m=0$) for this matrix element as
\begin{align}
\canon{\up{x}\Phi}{\hat{v}}{\up{w}\Phi_i^a} = \up{xw}\tilde{S} \qty(V_0 \up{ww}X_{ia} + 2\, \up{ww}[X(J-K)Y]_{ia}).
\end{align}
Distributing the zero-overlap orbitals among each contraction then leads to explicit
expressions for all $m$ values as
\begin{align}
\begin{split}
\braket{\up{x}\Phi}{&\hat{v}}{\up{w}\Phi_i^a} = 
\\
&
\begin{cases}
\up{xw}\tilde{S} \qty(V_0 \up{ww}X_{ia} + 2\, \up{ww}[X(J-K)Y]_{ia})
& m=0
\\[5pt]
\up{xw}\tilde{S} \qty(V_0 \up{ww}\bX_{ia} + 2\, \up{ww}[\bX(J-K)Y]_{ia} +\bar{V}_0 \up{ww}X_{ia} + 2\, \up{ww}[X(\bJ-\bK)Y]_{ia} + 2\, \up{ww}[X(J-K)\bY]_{ia})
& m=1
\\[5pt]
\up{xw}\tilde{S} \qty(\bar{\bar{V}}_0 \up{ww}X_{ia} + 2\, \up{ww}[\bX(\bJ-\bK)Y]_{ia} + \bar{V}_0 \up{ww}\bar{X}_{ia} + 2\, \up{ww}[\bX(J-K)\bY]_{ia} + 2\, \up{ww}[X(\bJ-\bK)\bY]_{ia})
& m=2
\\[5pt]
\up{xw}\tilde{S} \qty(\bar{\bar{V}}_0 \up{ww}\bX_{ia} + 2\, \up{ww}[\bX(\bJ-\bK)\bY]_{ia})
& m=3
\\[5pt]
0 & m>3.
\end{cases}
\end{split}
\end{align}
Next, we consider the two-body coupling of two singly excited determinants as 
\begin{align}
\braket{\up{x}\Phi_i^a}{\hat{v}}{\up{w}\Phi_{j}^{b}} 
= \sum_{pqrs} \up{x}v_{pqrs} 
\braket{\up{x}\Phi}{\up{w}b_i^{\dagger} \up{w}b_a^{\vphantom{\dagger}} \up{x}b_p^{\dagger} \up{x}b^{\dagger}_q \up{x}b_s \up{x}b_r   \up{w}b_b^{\dagger} \up{w}b_j^{\vphantom{\dagger}}  }{\up{w}\Phi}.
\end{align}
With a total of 24 possible ways to fully contract the corresponding matrix element, 
we maintain brevity by directly presenting the canonical form for $m=0$ as 
\begin{align}
\begin{split}
\canon{\up{x}\Phi_i^a}{\hat{v}}{\up{w}\Phi_{j}^{b}} 
=
\up{xw}\tilde{S} \times \Big[ & V_0 (\up{xx}X_{ai} \up{ww}X_{jb} + \up{wx}X_{ji} \up{xw}Y_{ab}  ) 
\\
&+2\qty(\up{xx}[Y(J-K)X]_{ai} \up{ww}X_{jb} + \up{ww}[X(J-K)Y]_{jb} \up{w}X_{ai}) 
\\
&+2\qty(\up{xw}[Y(J-K)Y]_{ab} \up{wx}X_{ji} - \up{wx}[X(J-K)X]_{ji} \up{xw}X_{ab} )
\\
&+2 \sum_{pqrs} \up{x}v_{pqrs} \up{xx}Y_{ap} \up{wx}X_{jq} (\up{xw}Y_{sb} \up{xx}X_{ri} -\up{xw}Y_{rb}  \up{xxw}X_{si} ) \Big].
\end{split}
\label{eq:sing_sing_two}
\end{align}
From here, expressions for the cases with $m>0$ can be recovered by distributing the zero-overlap orbitals
over the contractions associated with the $V$, $X$, $(J-K)$, or $Y$ matrices.
While partially contracted intermediate expressions could also be evaluated to avoid the nested summation
on the last line in Eq.~\eqref{eq:sing_sing_two}, this will generally incur an unacceptably large storage cost.

Finally, we consider the two-body coupling of a reference determinant and a double excitation 
\begin{align}
\braket{\up{x}\Phi}{\hat{v}}{\up{w}\Phi_{ij}^{ab}} 
= \sum_{pqrs} \up{x}v_{pqrs} 
\braket{\up{x}\Phi}{ \up{x}b_p^{\dagger} \up{x}b^{\dagger}_q \up{x}b_s \up{x}b_r  \up{w}b_a^{\dagger} \up{w}b_b^{\dagger} \up{w}b_j^{\vphantom{\dagger}} \up{w}b_i^{\vphantom{\dagger}} }{\up{w}\Phi}.
\end{align}
Again, there are a total of 24 possible ways to fully contract the corresponding matrix element, 
so we advance directly to the canonical form for $m=0$, given as
\begin{align}
\begin{split}
\canon{\up{x}\Phi}{\hat{v}}{\up{w}\Phi_{ij}^{ab}} 
=
\up{xw}\tilde{S} \times \Big[ & V_0 (\up{ww}X_{ia} \up{ww}X_{jb} - \up{ww}X_{ib} \up{ww}X_{ja}  ) 
\\
&+2\qty(\up{ww}[X(J-K)Y]_{ia} \up{ww}X_{jb} - \up{ww}[X(J-K)Y]_{ib} \up{ww}X_{ja}) 
\\
&+2\qty(\up{ww}[X(J-K)Y]_{jb} \up{ww}X_{ia} - \up{ww}[X(J-K)Y]_{ja} \up{ww}X_{ib} )
\\
&+2 \sum_{pqrs} \up{x}v_{pqrs} \up{wx}X_{ip} \up{wx}X_{jq} (\up{xw}Y_{sb} \up{xw}Y_{ra} - \up{xw}Y_{sa} \up{xw}Y_{rb}) \Big].
\end{split}
\label{eq:ref_doub_two}
\end{align}
We note that this formula has previously been identified  in Ref.~\onlinecite{Nite2019b} for $m=0$, but
our approach now allows this result to be extended for $m>0$.
 
\end{widetext}

\subsection{Illustration of Scaling}
\label{subsec:AnalysisOfScaling}

\hugh{Through the use of intermediate matrices in Sections~\ref{subsec:AsymmRepresentation}--\ref{subsec:OneBody},
the cost of evaluating overlap and one-body matrix elements between excited configurations can be made
independent of the number of electrons $\Ne$ or basis functions $\Nbas$.
In both cases, the computational cost of evaluating the intermediate matrices scales as $\Or{\Nref^2\, \Nbas^3}$, which is 
determined by the cost of evaluating the screened overlap terms in Eqs.~\eqref{eq:funContractLeft} 
and \eqref{eq:funContractRight}.
The subsequent cost of evaluating matrix elements between excited configurations then scales as $\Or{1}$.
In contrast, the cost of applying the generalised Slater--Condon rules for overlap or one-body operators
is dominated by the evaluation of the occupied orbital overlap matrix, giving a scaling of $\Or{\Nbas^2 \Ne}$
for \emph{every} excited coupling term.}

\hugh{
For two-body operators, a similar application of two-body intermediates requires transformations of the two-electron
integrals into an asymmetric MO representation for each pair of determinants, and for every possible combination of
four contractions in the four possible forms given in 
Eq.~\eqref{eq:funPairContract} or \eqref{eq:funPairContract2}.
These two-body intermediates therefore carry a storage overhead that scales as $\Or{4^4 \Nref^2\,\Nbas^4}$, which 
will generally be unacceptably large. 
We imagine there may be certain applications that require only a subset of these two-electron intermediates, 
which could then be stored to achieve the subsequent $\Or{1}$ scaling. 
In general, however, the $\Nbas^4$ scaling for two-body matrix elements cannot overcome, although the overall
cost is still reduced by avoiding biorthogonalisation of the occupied orbitals in 
the excited configurations.}

\hugh{To explicitly illustrate the computational speed-up that might be expected using the extended nonorthogonal Wick's 
theorem, consider the evaluation of matrix elements between two configuration interaction singles (CIS)
wave functions built from different reference Slater determinants
\begin{subequations}
\begin{align}
\ket{\up{x}\Psi_\text{CIS}} &= \sum_{ia} \up{x}t_{ia} \ket{\up{x}\Phi_{ia}}
\\
\ket{\up{w}\Psi_\text{CIS}} &= \sum_{jb} \up{w}t_{jb} \ket{\up{w}\Phi_{jb}}.
\end{align}
\end{subequations}
In particular, consider the evaluation of the overlap 
\begin{equation}
\brkt{\up{x}\Psi_\text{CIS}}{\up{w}\Psi_\text{CIS}} 
= 
\sum_{ia,jb} \up{x}t^{*}_{ia}\, \up{w}t^{\vphantom{*}}_{jb} \brkt{\up{x}\Phi_{ia}}{\up{w}\Phi_{jb}},
\end{equation}
a one-body operator $\hat{f}$, e.g.\ the transition dipole moment, 
\begin{equation}
\braket{\up{x}\Psi_\text{CIS}}{\hat{f}}{\up{w}\Psi_\text{CIS}} 
= 
\sum_{ia,jb} \up{x}t^{*}_{ia}\, \up{w}t^{\vphantom{*}}_{jb} \braket{\up{x}\Phi_{ia}}{\hat{f}}{\up{w}\Phi_{jb}},
\end{equation}
and a two-body operator $\hat{v}$ such as the two-electron repulsion
\begin{equation}
\braket{\up{x}\Psi_\text{CIS}}{\hat{v}}{\up{w}\Psi_\text{CIS}} 
= 
\sum_{ia,jb} \up{x}t^{*}_{ia}\, \up{w}t^{\vphantom{*}}_{jb} \braket{\up{x}\Phi_{ia}}{\hat{v}}{\up{w}\Phi_{jb}}.
\end{equation}
The relevant overlap, one-body, and two-body nonorthogonal matrix elements are given in Eqs~\eqref{eq:sing_sing_ov},
\eqref{eq:fockSingSing}, and \eqref{eq:sing_sing_two} respectively. 
In each case, only one set of intermediates need to be evaluated using the biorthogonalised reference determinants, 
followed by a total of $\Ne^2 \Nbas^2$ nonorthogonal matrix elements between singly excited configurations
(assuming $\Nbas\gg \Ne$).
Here, we only consider one pairing of the reference determinants.}

\hugh{%
The computational scaling associated with first evaluating the intermediate terms from the reference determinants, and 
then evaluating all the nonorthogonal coupling terms between the excited configurations is summarised in Table~\ref{tab:CISscaling}.
For the overlap and one-body terms, the computational scaling is dominated by the evaluation of the intermediates, 
which scales as $\Or{\Nbas^3}$.
In comparison, the total cost of biorthogonalising the occupied orbitals for every pair of excited configurations
scales as $\Or{\Ne^3 \Nbas^4}$.
For two-body operators, the cost for both approaches is dominated by the contraction with the two-body 
integrals for each pair of excited configurations. 
This leads to a total scaling $\Or{\Ne^2 \Nbas^6}$ for both cases, but the extended nonorthogonal Wick's theorem may 
still be faster as the prefactor is reduced for each element.
}

\begin{table}[th]
\caption{
Computational scaling for evaluating the overlap, one-body, and two-body coupling 
 between two CIS wave functions with different reference determinants
using the generalised Slater--Condon rules (Slat.--Con.) or extended non-orthogonal Wick's theorem.
\label{tab:CISscaling}}
\begin{ruledtabular}
\begin{tabular}{lcccc}
        & \mc{2}{c}{Intermediate}  &	\mc{2}{c}{Subsequent Cost} \\
                   \cline{2-3}	                       \cline{4-5} 
Coupling Term    &  Slat.--Con.	&  Wick's       & Slat.--Con. & Wick's \\
\hline
$\brkt{\up{x}\Psi_\text{CIS}}{\up{w}\Psi_\text{CIS}}$ 
        &	N/A	        & $\Nbas^3$        &  $\Ne^3 \Nbas^4$   &  $\Ne^2 \Nbas^2$ \\
$\braket{\up{x}\Psi_\text{CIS}}{\hat{f}}{\up{w}\Psi_\text{CIS}}$
        &   N/A           & $\Nbas^3$        &  $\Ne^3 \Nbas^4$   &  $\Ne^2 \Nbas^2$ \\
$\braket{\up{x}\Psi_\text{CIS}}{\hat{v}}{\up{w}\Psi_\text{CIS}}$
        &   N/A           & $\Nbas^3$        &  $\Ne^2 \Nbas^6$   &  $\Ne^2 \Nbas^6$ \\
\end{tabular}
\end{ruledtabular}
\end{table}

\section{Concluding Remarks}
\label{sec:ConcludingRemarks}

Intuitive derivations and efficient implementations of nonorthogonal matrix elements
are becoming increasingly important for the development of nonorthogonal configuration interaction
methods and inter-state coupling terms for state-specific excited state wave functions.
However, until now, the evaluation of these matrix elements has relied on the generalised 
Slater--Condon rules, while the second-quantised nonorthogonal Wick's theorem fails when there 
are any zero-overlap orbital pairs between the reference determinants. 
In this work, we have extended the nonorthogonal Wick's theorem to the case where 
two determinants have nonorthogonal orbitals, but have a zero many-electron overlap.
This new theory, which we call the Extended Nonorthogonal Wick's theorem, provides the most
generalised framework for evaluating matrix elements between two nonorthogonal determinants
using second quantisation.

Among the primary advantages of our new approach is the ease of deriving and evaluating 
matrix elements between excited configurations from nonorthogonal reference determinants.
To illustrate this feature, we have derived a series of overlap, one-body, and two-body coupling 
terms between nonorthogonal excited configurations.
For overlap terms and one-body operators, these excited nonorthogonal matrix elements can be 
expressed in terms of one-body intermediate terms that can be precomputed and stored 
for a given pair of reference determinants. 
As a result, the subsequent cost of evaluating the coupling of excited configurations scales 
\hugh{as $\Or{1}$, which is the same as the conventional Slater--Condon rules or Wick's theorem
for orthogonal reference determinants.
However, in the current form of the theory, the cost of evaluating two-body coupling terms scales as $\Or{\Nbas^4}$,
which is the same as applying the generalised Slater--Condon rules for each pair of nonorthogonal 
excited configurations.
From a computational perspective, the extended nonorthogonal Wick's theorem will therefore provide
the greatest acceleration in tasks that require at most one-body coupling terms, such as the evaluation
of transition dipole moments for orbital-optimised excited states,\cite{Shea2018,Hardikar2020,Tran2019,Tran2020} 
or the coupling elements of 
a one-body reference Hamiltonian in nonorthogonal perturbation theories.\cite{Yost2013,Yost2016,Yost2019,Burton2020}}

\hugh{Looking forwards, we hope that this work will encourage
and accelerate future developments in nonorthogonal electronic structure theory by creating a unifying
theory for deriving challenging nonorthogonal matrix elements.
As part of this vision, we are working on an open-source C++ library for evaluating typical 
nonorthogonal matrix elements, and we intend to report on this project soon.  
}

\section*{Acknowledgements}
HGAB was support by New College, Oxford through the Astor Junior Research Fellowship.
The author is grateful to Rebecca Lloyd for careful proof-reading.
\section*{Data Availability Statement}
Data sharing not applicable -- no new data generated.

\appendix
\section{Thouless' Theorem}
\label{apdx:ThoulessThereom}
Thouless' Theorem\cite{Thouless1960} allows any Slater determinant to be represented using only 
single excitations from another determinant, i.e.\
\begin{align}
\ket{\up{w}\Phi} = \exp(\cZ) \ket{\up{x}\Phi}.
\label{eq:thouless}
\end{align}
Here, we follow Ref.~\onlinecite{Jimenez-Hoyos2012a} and provide a brief derivation of this theorem.
First, the two sets of second-quantisation operators can be related as
\begin{subequations}
	\begin{align}
	\up{w}\bdag_{i} 
	&= 
    \sum_j \up{x}\bdag_{j}\, \up{xw}A_{ji}  + \sum_b \up{x}\bdag_{b}\, \up{xw}B_{bi}
	\\
    \hugh{\up{w}\bdag_{a}}
	&= 
    \sum_j \up{x}\bdag_{j}\, \up{xw}\tilde{B}_{ja} + \sum_b \up{x}\bdag_{b}\, \up{xw}D_{ba},
	\end{align}	                         
\end{subequations}
where 
\begin{subequations}
\begin{align}
\up{xw}A_{ji} &= \sum_{\mu \nu} (^{x}C^{*})^{\cdot \nu}_{j \cdot} g_{\nu \mu} (^{w} C)^{\mu \cdot}_{\cdot i}
\\
\up{xw}B_{bi} &=\sum_{\mu \nu} (^{x}C^{*})^{\cdot \nu}_{b \cdot} g_{\nu \mu} (^{w} C)^{\mu \cdot}_{\cdot i}
\\ 
\up{xw}\tilde{B}_{ja} &= \sum_{\mu \nu} (^{x}C^{*})^{\cdot \nu}_{j \cdot} g_{\nu \mu} (^{w} C)^{\mu \cdot}_{\cdot a}
\\
\up{xw}D_{ba} &= \sum_{\mu \nu} (^{x}C^{*})^{\cdot \nu}_{b \cdot} g_{\nu \mu} (^{w} C)^{\mu \cdot}_{\cdot a}
\end{align}
\end{subequations}
are sub-blocks of the orbital overlap matrix.
Taking a suitable transformation among the orbitals of $\ket{\Phi_{w}}$ allows these to be further reduced to 
\begin{subequations}
	\begin{align}
	\up{w}\tbdag_{i} &= \up{x}\bdag_{i} + \sum_{ja} \up{x}\bdag_{a}\, \up{xw}B_{aj} (^{xw} A^{-1} )_{ji} 
	\\
	\up{w}\tbdag_{a} &= \up{x}\bdag_{a} + \sum_{jb} \up{x}\bdag_{j}\, \up{xw}\tilde{B}_{jb} (^{xw}D^{-1})_{ba}.
	\end{align}	                         
\end{subequations}
The occupied orbitals can then be represented by the transformation
	\begin{align}
	\up{w}\tbdag_{i} &= \up{x}\bdag_{i} + \sum_a  \up{x}\bdag_{a}\, \up{xw}Z_{ai}
	\end{align}
where now 
\begin{align}
	\up{xw}Z_{ai} = \sum_{j} \up{xw}B_{aj} (^{xw} A^{-1} )_{ji}.
\end{align}
The overall transformation in Eq.~\eqref{eq:thouless} is then given by 
\begin{align}
\cZ = \up{xw}Z_{ai} \up{x}\bdag_{a} \up{x}\b_{i}.
\end{align}
Finally, if the occupied orbitals  are represented in a biorthogonal basis satisfying Eq.~\eqref{eq:LowdinPairing}, 
then the $Z$ matrix elements are given simply as
\begin{align}
\up{xw}Z_{ai} = \sum_{\mu \nu} (^{x}\tC^{*})^{\cdot \mu}_{a \cdot}\, g_{\mu \nu}\, (^{w}\tC)^{\nu \cdot}_{\cdot i} \frac{1}{\up{xw}\tilde{S}_{i}}.
\end{align}

\section{Similarity Transformed Operators}
\label{apdx:SimTransform}
To evaluate common matrix elements using the single excitation operators defined in Eq.~\eqref{eq:zero_transformed_excitation},
we follow Ref.~\onlinecite{Jimenez-Hoyos2012a} and consider the similarity transformed operators
\begin{subequations}
	\begin{align}
(d[\mathsmaller{\Cn}]^{\mu})^{\dagger}  &= \exp \big(-\tilde{\cZ}^{\cC_n} \big) (a^{\mu})^{\dagger} \exp \big(\tilde{\cZ}^{\cC_n} \big)
\\
d[\mathsmaller{\Cn}]^{\mu} &= \exp \big(-\tilde{\cZ}^{\cC_n} \big) a^{\mu} \exp \big(\tilde{\cZ}^{\cC_n} \big).
	\end{align}
\end{subequations}
First, consider the expansion of $(d[\Cn]^{\mu})^{\dagger}$ as
\begin{align}
\begin{split}
(d[\mathsmaller{\Cn}]^{\mu})^{\dagger} 
&= \exp(-\tilde{\cZ}^{\Cn} ) (a^{\mu})^{\dagger} \exp(\tilde{\cZ}^{\Cn} )
\\
&=  (a^{\mu})^{\dagger} - [\tilde{\cZ}^{\Cn} ,  (a^{\mu})^{\dagger}].
\end{split}
\end{align}
\hugh{Expanding $(a^{\mu})^{\dagger}$ using Eq.~\eqref{eq:AOcontra} and inserting the definition
of $\tilde{\cZ}^{\Cn}$ from Eq.~\eqref{eq:zero_transformed_excitation} leads to} 
\begin{widetext}
\hugh{
\begin{align}
\begin{split}
(d[\mathsmaller{\Cn}]^{\mu})^{\dagger} 
&= 
\sum_p \up{x}\tbdag_p\, 
(\up{x}\tC^{*})^{\cdot \mu}_{p \cdot} 
- \sum_{ap} \Bigg(
\sum_{\{i | \up{xw}\tilde{S}_{i} \neq 0 \} } \up{xw}Z_{ai}[\up{x}\tbdag_a \up{x}\tb_i, \up{x}\tbdag_p ] 
+
\sum_{k \in \mathsmaller{\Cn}} \up{xw}\tilde{S}_{ak} [\up{x}\tbdag_a \up{x}\tb_k, \up{x}\tbdag_p ] 
\Bigg) 
(\up{x}\tC^{*})^{\cdot \mu}_{p \cdot}. 
\end{split}
\end{align}
}

Noting the commutation relation
$%
[\up{x}\tbdag_{a} \up{x}\tb_{i}, \up{x}\tbdag_{p}] = \up{x}\tbdag_{a} \delta_{ip}$,
\hugh{then allows this expression to be simplified as}
\begin{align}
\begin{split}
(d[\mathsmaller{\Cn}]^{\mu})^{\dagger} 
= 
\sum_{p}
\up{x}\tbdag_{p} (^{x}\tC^{*})^{\cdot \mu}_{p \cdot}
- \sum_{a} \up{x}\tbdag_{a} \qty( \sum_{\{i | \up{xw}\tilde{S}_{i} \neq 0 \} } \hspace{-5pt}
 \up{xw}Z_{ai}   (^{x}\tC^{*})^{\cdot \mu}_{i \cdot}   + \sum_{k \in \Cn}  \up{xw}\tS_{ak} (^{x}\tC^{*})^{\cdot \mu}_{k\cdot}  ).
\end{split}
\end{align}
Finally, separating the summation over $p$ into a summation over the occupied orbitals $i$ and
the virtual orbitals $a$ recovers the form given in Eq.~\eqref{eq:TransformedOperators2a}.
Applying a similar approach for the $(d[\Cn]^{\mu})$ operators then recovers 
the expression in Eq.~\eqref{eq:TransformedOperators2b}.

\section{Similarity Transformed Contractions}
\label{apdx:SimTransformContract}
\hugh{
To evaluate the contractions $\wick{(\c1 d[\mathsmaller{\Cn}]^{\mu})^{\dagger} (\c1 d[\mathsmaller{\Cn}]^{\nu})}$ and 
$\wick{(\c1 d[\mathsmaller{\Cn}]^{\mu}) (\c1 d[\mathsmaller{\Cn}]^{\nu})^{\dagger}}$ with respect to the Fermi
vacuum $\braket{\up{x}\Phi}{\cdots}{\up{x}\Phi}$, we first expand the transformed operators in terms of the 
$\up{x}\tb_p$ and $\up{x}\tbdag_p$ operators. 
The only non-zero contractions between these MO operators are 
$\wick{\up{x}\c1 \tb_{a} \up{x}\c1\tb^{\dagger}_{b} } = \delta_{ab}$ 
and $\wick{\up{x}\c1 \tb^{\dagger}_{i} \up{x}\c1 \tb_{j}} = \delta_{ij}$.
We can therefore expand the contracted product
$\wick{(\c1 d[\mathsmaller{\Cn}]^{\mu})^{\dagger} (\c1 d[\mathsmaller{\Cn}]^{\nu})}$ as 
\begin{align}
\wick{(\c1 d[\mathsmaller{\Cn}]^{\mu})^{\dagger} (\c1 d[\mathsmaller{\Cn}]^{\nu})} 
&=
\sum_{i} (^{x}\tC)^{\nu \cdot}_{\cdot i} (^{x}\tC^{*})^{\cdot \mu}_{i \cdot}
+
\sum_{a} \qty( \sum_{ \{i | \up{xw}\tS_i \neq 0 \} }\hspace{-5pt} (^{x}\tC)^{\nu \cdot}_{\cdot a}\, \up{xw}Z_{ai} (^{x}\tC^{*})^{\cdot \mu}_{i \cdot}
+
 \sum_{k \in \cC_n} (^{x}\tC)^{\nu \cdot}_{\cdot a}\, \up{xw}\tS_{ak} (^{x}\tC^{*})^{\cdot \mu}_{k \cdot}).
\end{align}
Inserting the definition of $\up{xw}Z_{ai}$ and $\up{xw}\tS_{ak}$ from Eqs.~\eqref{eq:Zdef} and \eqref{eq:PairedOccVir}
respectively then leads to the form
\begin{align}
\wick{(\c1 d[\mathsmaller{\Cn}]^{\mu})^{\dagger} (\c1 d[\mathsmaller{\Cn}]^{\nu})} 
&=
\sum_{i} (^{x}\tC)^{\nu \cdot}_{\cdot i} (^{x}\tC^{*})^{\cdot \mu}_{i \cdot}
+
\sum_{a}
\sum_{\sigma \tau}
 (^{x}\tC)^{\nu \cdot}_{\cdot a}\,   (^{x}\tC^{*})^{\cdot \sigma}_{a \cdot}\, g_{\sigma \tau}
\qty(
\sum_{ \{i | \up{xw}\tS_i \neq 0 \} }\hspace{-5pt}
\, (^{w}\tC)^{\tau \cdot}_{\cdot i} \frac{1}{\up{xw}\tS_{i}}  
(^{x}\tC^{*})^{\cdot \mu}_{i \cdot}
+
\sum_{k \in \cC_n}
(\up{w}\tC)^{\tau \cdot}_{\cdot k} (^{x}\tC^{*})^{\cdot \mu}_{k \cdot}).
\label{eq:ExpandedForm}
\end{align}
Resolving the identity allows us to obtain the additional relationship
\begin{align}
\sum_{a} (^{x}\tC)^{\nu \cdot}_{\cdot a}\,   (^{x}\tC^{*})^{\cdot \sigma}_{a \cdot}
=
g^{\nu \sigma}
-
\sum_{j} (^{x}\tC)^{\nu \cdot}_{\cdot j}\,   (^{x}\tC^{*})^{\cdot \sigma}_{j \cdot}
\end{align}
which, when inserted into Eq.~\eqref{eq:ExpandedForm}, leads to 
\begin{align}
\begin{split}
\wick{(\c1 d[\mathsmaller{\Cn}]^{\mu})^{\dagger} (\c1 d[\mathsmaller{\Cn}]^{\nu})} 
=
\sum_{ \{k | \up{xw}\tS_k = 0 \} }
(^{x}\tC)^{\nu \cdot}_{\cdot k} (^{x}\tC^{*})^{\cdot \mu}_{k \cdot}
+
\sum_{ \{i | \up{xw}\tS_i \neq 0 \} }\hspace{-5pt}
\, (^{w}\tC)^{\nu \cdot}_{\cdot i} \frac{1}{\up{xw}\tS_{i}}  
(^{x}\tC^{*})^{\cdot \mu}_{i \cdot}
+
\sum_{k \in \cC_n}
(\up{w}\tC)^{\nu \cdot}_{\cdot k} (^{x}\tC^{*})^{\cdot \mu}_{k \cdot}
\end{split}
\label{eq:ExpandedForm2}
\end{align}
To obtain Eq.~\eqref{eq:ExpandedForm2} from Eq.~\eqref{eq:ExpandedForm}, we have exploited the relationships
\begin{align}
\sum_{ \{i | \up{xw}\tS_i \neq 0 \} }
\sum_{j} 
\sum_{\sigma \tau}
(^{x}\tC)^{\nu \cdot}_{\cdot j}
\,   (^{x}\tC^{*})^{\cdot \sigma}_{j \cdot}\, g_{\sigma \tau}
\, (^{w}\tC)^{\tau \cdot}_{\cdot i} \frac{1}{\up{xw}\tS_{i}}  
(^{x}\tC^{*})^{\cdot \mu}_{i \cdot}
= 
\sum_{ \{i | \up{xw}\tS_i \neq 0 \} }
(\up{x}\tC)^{\nu \cdot}_{\cdot i} (\up{x}\tC^{*})^{\cdot \mu}_{i \cdot}
\end{align}
and
\begin{align}
\sum_{ k \in \mathsmaller{\Cn} }
\sum_{j} 
\sum_{\sigma \tau}
(^{x}\tC)^{\nu \cdot}_{\cdot j}\,   
(^{x}\tC^{*})^{\cdot \sigma}_{j \cdot}\, g_{\sigma \tau}
\, (^{w}\tC)^{\tau \cdot}_{\cdot k}
(^{x}\tC^{*})^{\cdot \mu}_{k \cdot}
= 
0
\end{align}
Finally, Eq.~\eqref{eq:LeftContraction} is recovered by modifying the separation of summation indices
in Eq.~\eqref{eq:ExpandedForm2} to give the form
\begin{align}
\begin{split}
\wick{(\c1 d[\mathsmaller{\Cn}]^{\mu})^{\dagger} (\c1 d[\mathsmaller{\Cn}]^{\nu})} 
&=
\\
\hspace{-5pt}
\sum_{ \{k | \up{xw}\tS_k = 0 \} }
\hspace{-5pt}
&(^{x}\tC)^{\nu \cdot}_{\cdot k} (^{x}\tC^{*})^{\cdot \mu}_{k \cdot}
+
\hspace{-5pt}
\sum_{ \{k | \up{xw}\tS_k = 0 \} }
\hspace{-5pt}
(\up{w}\tC)^{\nu \cdot}_{\cdot k} (^{x}\tC^{*})^{\cdot \mu}_{k \cdot}
+
\sum_{ \{i | \up{xw}\tS_i \neq 0 \} }\hspace{-5pt}
\, (^{w}\tC)^{\nu \cdot}_{\cdot i} \frac{1}{\up{xw}\tS_{i}}  
(^{x}\tC^{*})^{\cdot \mu}_{i \cdot}
-
\sum_{k \notin \cC_n}
(\up{w}\tC)^{\nu \cdot}_{\cdot k} (^{x}\tC^{*})^{\cdot \mu}_{k \cdot}
\end{split}
\end{align}
and introducing the matrices defined in Eqs.~\eqref{eq:CoDensity}--\eqref{eq:Msum}.
The second contraction, Eq.~\eqref{eq:RightContraction}, then follows from 
\begin{align}
\wick{(\c1 d[\mathsmaller{\Cn}]^{\mu}) (\c1 d[\mathsmaller{\Cn}]^{\nu})^{\dagger}} 
= g^{\mu \nu} - \wick{(\c1 d[\mathsmaller{\Cn}]^{\nu})^{\dagger} (\c1 d[\mathsmaller{\Cn}]^{\mu})}. 
\end{align}
}
\end{widetext}

\section*{References}
\bibliography{manuscript}

\end{document}